\newif\iftechreport%
\techreporttrue%

\RequirePackage{etoolbox}
\ifdef{\iftechreport}{}{\newif\iftechreport}
\documentclass[conference]{IEEEtran}
\IEEEoverridecommandlockouts

\usepackage{tikz}
\usepackage{amsmath}
\usepackage{filecontents}
\usepackage[hidelinks]{hyperref}
\usepackage{placeins}
\usepackage[export]{adjustbox}
\usepackage{url}
\usepackage{booktabs}
\usepackage{tabularx}
\usepackage{threeparttable}

\usepackage{siunitx}
\usepackage{multirow}
\usepackage{graphicx}
\usepackage{setspace}
\usepackage{float}
\usepackage{amsthm}
\usepackage{thmtools,thm-restate}
\declaretheoremstyle[
spaceabove=0pt,
spacebelow=0pt,
headindent=\parindent,
numbered=no,
headfont=\normalfont\bfseries,
bodyfont=\normalfont\itshape,
]{examplestyle}
\declaretheorem[style=examplestyle]{example}

\newtheorem{theorem}{Theorem}

\usepackage[ruled,vlined,linesnumbered]{algorithm2e}
\SetInd{0.5em}{0.3em}
\SetKwProg{Fn}{Function}{}{}
\SetKwRepeat{Do}{do}{while}
\SetVlineSkip{0pt}

\SetCommentSty{mycommfont}
\SetAlCapNameFnt{\small}
\SetAlCapFnt{\small}
\SetAlFnt{\small}
\makeatletter
\patchcmd\algocf@Vline{\vrule}{\vrule \kern-0.4pt}{}{}
\patchcmd\algocf@Vsline{\vrule}{\vrule \kern-0.4pt}{}{}
\newcommand{\removelatexerror}{\let\@latex@error\@gobble}
\makeatother
\usepackage{amsmath}
\usepackage{balance}
\usepackage{graphicx}
\usepackage[inline]{enumitem}
\setlist[itemize]{noitemsep,partopsep=0pt,topsep=0pt,leftmargin=1.5em}
\usepackage{mathtools}
\usepackage{verbatim}
\usepackage{threeparttable}
\usepackage{footnote}
\makesavenoteenv{tabular}
\makesavenoteenv{table}
\DeclarePairedDelimiter{\ceil}{\lceil}{\rceil}

\usepackage[capitalise, noabbrev]{cleveref}
\usepackage{xcolor}
\usepackage{pifont}
\newcommand{\eat}[1]{}
\newenvironment{myproof}{\renewcommand{\proofname}{Proof}\proof}{\endproof}

\apptocmd\normalsize{%
	\setlength{\abovedisplayskip}{1pt}
	\setlength{\belowdisplayskip}{1pt}
	\setlength{\abovedisplayshortskip}{1pt}
	\setlength{\belowdisplayshortskip}{1pt}
}{}{}

\newcommand{\ce}[1]{{\color{black}{#1}}}
\DeclareMathOperator{\memindex}{RS-tree}
\DeclareMathOperator{\memindexs}{RS-trees}
\newcommand\rev[1]{\textcolor{black}{#1}}

\AtBeginDocument{%
	\providecommand\BibTeX{{%
			Bib\TeX}}}

\def\BibTeX{{\rm B\kern-.05em{\sc i\kern-.025em b}\kern-.08em
    T\kern-.1667em\lower.7ex\hbox{E}\kern-.125emX}}

\begin{document}
	
	\iftechreport
	\title{COLE$^+$: Towards Practical Column-based Learned Storage for Blockchain Systems }
	\else
	\title{COLE$^+$: Towards Practical  Column-based Learned Storage for Blockchain Systems}
	\fi 

\author{%
{Ce Zhang{\small$^{*}$}, Cheng Xu{\small$^{*}$}, Haibo Hu{\small$^{\ddag}$}, Jianliang Xu{\small$^{*}$}}%
\vspace{1.6mm}\\
\fontsize{10}{10}\selectfont\itshape%
$^{*}$\,Department of Computer Science, Hong Kong Baptist University, Hong Kong SAR\\
$^{\ddag}$\,Department of Electrical and Electronic Engineering, Hong Kong Polytechnic University, Hong Kong SAR \\
\fontsize{9}{9}\selectfont\ttfamily\upshape%
\{cezhang, chengxu, xujl\}@comp.hkbu.edu.hk, haibo.hu@polyu.edu.hk \\
}
	
	

	\maketitle

    \begin{abstract}
		Blockchain provides a decentralized and tamper-resistant ledger for securely recording transactions across a network of untrusted nodes. While its transparency and integrity are beneficial, the substantial storage requirements for maintaining a complete transaction history present significant challenges. For example, Ethereum nodes require around 23TB of storage, with an annual growth rate of 4TB. Prior studies have employed various strategies to mitigate the storage challenges. Notably, COLE significantly reduces storage size and improves throughput by adopting a column-based design that incorporates a learned index, effectively eliminating data duplication in the storage layer. However, this approach has limitations in supporting chain reorganization during blockchain forks and state pruning to minimize storage overhead. In this paper, we propose COLE$^+$, an enhanced storage solution designed to address these limitations. COLE$^+$ incorporates a novel rewind‐supported in-memory tree structure for handling chain reorganization, leveraging content‐defined chunking (CDC) to maintain a consistent hash digest for each block. For on‐disk storage, a new two-level Merkle Hash Tree (MHT) structure, called prunable version tree, is developed to facilitate efficient state pruning. Both theoretical and empirical analyses show the effectiveness of COLE$^+$ and its potential for practical application in real-world blockchain systems. 
	\end{abstract}
	
	\section{Introduction}\label{sec:intro}
	Blockchain, a decentralized and append-only ledger, leverages cryptographic hash chains and distributed consensus protocols to securely record transactions across a network of untrusted nodes~\cite{nakamoto2008bitcoin, wood2014ethereum}. Its inherent transparency and tamper-resistance have established it as a foundational technology for cryptocurrencies and a wide range of decentralized applications. \rev{To ensure complete and verifiable data provenance, blockchain nodes must maintain a comprehensive history of transactions and ledger states, enabling users to query historical data with strong integrity guarantees.} However, this comprehensive record-keeping incurs substantial storage overhead, which grows rapidly as the blockchain expands. For example, as of October 2025, Ethereum nodes require approximately 23 TB of storage, with an annual growth rate of 
    $\sim$4 TB~\cite{ethsize}.
	
\begin{figure}[t]
		\centering
		\includegraphics[width=.85\linewidth]{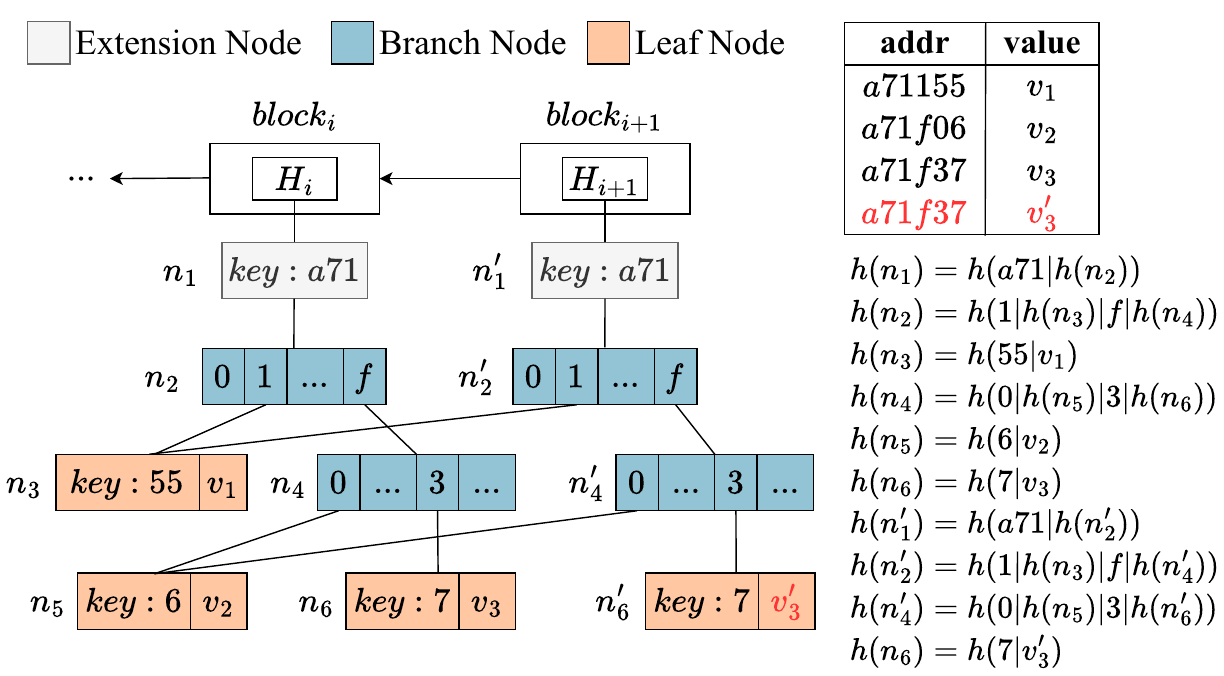}
        \vspace{-1em}
		\caption{An Example of Merkle Patricia Trie}\label{fig:mpt}
	\end{figure}

	Prior studies~\cite{tian2024letus, zhang2024cole} show that the excessive storage overhead stems from the underlying index, Merkle Patricia Trie (MPT)~\cite{wood2014ethereum}. MPT retains obsolete nodes during data updates to enable data provenance via pointer chasing. For example, as shown in \cref{fig:mpt}, updating address $a71f37$ with value $v_3'$ in block $i+1$ adds a new path ($n_1'$, $n_2'$, $n_4'$, $n_6'$) for the updated value while retaining the existing path ($n_1$, $n_2$, $n_4$, $n_6$) for the historical value $v_3$. This design allows access to any historical value by traversing the index nodes under a specific historical block (e.g., traversing $n_1$, $n_2$, $n_4$, $n_6$ retrieves $v_3$ in block $i$). However, it also introduces significant storage overhead due to duplication along the updated path (e.g., $n_1'$, $n_2'$, $n_4'$, $n_6'$, and $n_1$, $n_2$, $n_4$, $n_6$).
	
	Existing work has explored various strategies to address the storage challenges posed by MPT. LETUS~\cite{tian2024letus} uses delta-encoding to reduce the storage requirements of the blockchain index. It also adopts page-based, simple file storage instead of key-value databases (e.g., RocksDB~\cite{rocksdb}) to enable fine-grained I/O optimizations. Compared with MPT, LETUS achieves up to 4$\times$ storage reduction, as reported in \cite{tian2024letus}. However, its proposed index, DMM-Tree, still employs an MPT-like node duplication strategy during data updates, leaving room for further storage optimization.
	SlimArchive~\cite{feng2024slimarchive} opts to eliminate the Merkle-based structures and flattens the minimal blockchain state changes. However, this approach no longer supports data authentication and provenance, which are critical security features of blockchain systems. 
	
\begin{figure}[t]
		\centering
		\includegraphics[width=.75\linewidth]{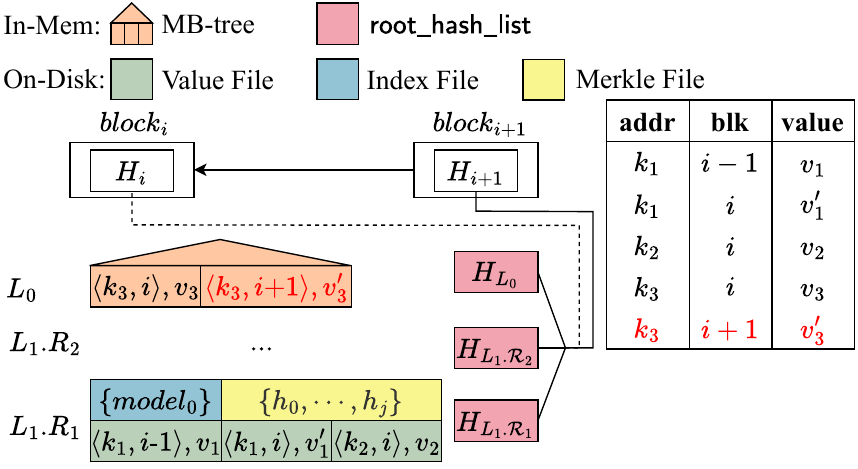}
        \vspace{-1em}
		\caption{Structure of COLE~\cite{zhang2024cole}}\label{fig:cole-overview}
	\end{figure}

	In contrast, COLE employs a column-based design coupled with a learned index to significantly reduce storage overhead (by up to 5.8$\times$ against MPT) while maintaining efficient data access~\cite{zhang2024cole}.
    \Cref{fig:cole-overview} illustrates the overall design of COLE, where a log-structured merge-tree (LSM-tree) consisting of multiple sorted runs is used to manage blockchain states. For each on-disk sorted run, COLE uses a value file to store the ledger states as a database column, an index file to predict the states' positions in the value file via a tailored learned index, and a Merkle file to ensure the data integrity of the value file. COLE's column-based design, in which successive versions of each state are stored contiguously, enables efficient data retrieval and reduces storage size. The learned index leverages the inherent ordering of the data for fast data access. \rev{At the same time, maintaining the historical states within the index of the  latest block eliminates MPT tree node duplication, thereby yielding substantial storage savings and enabling rapid historical state retrieval without the need to traverse prior block indexes.}
	
	Despite COLE's effectiveness in reducing storage size and enhancing system throughput, several limitations hinder its wide practical deployment:
	
	\begin{itemize}
		\item \textbf{Limited Chain Reorganization Support}: 
        \rev{COLE struggles to handle chain reorganizations during blockchain forks~\cite{chainreorg}. These forks can arise from either consensus protocols (e.g., Proof of Work~\cite{nakamoto2008bitcoin}) or software upgrades (e.g., post-Ethereum DAO attack hard fork~\cite{ethfork}).
        The former, which occurs frequently, typically rewinds a few recent blocks and thus requires COLE’s in-memory states to revert to a previous version.}
        However, this is infeasible because its underlying in-memory Merkle B-tree structure depends on both the data values and their insertion order~\cite{wang2018forkbase, yue2020analysis}. Rewinding blockchain states in this manner would introduce inconsistencies in the tree index across nodes and violate the requirement for a globally agreed-upon root hash. In the latter, rarer case, reversing the index is also difficult due to modifications in on-disk runs caused by flush and recursive merge operations in the LSM-tree. As a result, COLE is limited to blockchain systems that do not fork~\cite{gilad2017algorand, wood2016polkadot, androulaki2018hyperledger}.

		\item \textbf{Lack of State Pruning Support}: 
        \rev{COLE does not support state pruning, a common technique used by blockchain full nodes to reduce storage overhead by retaining only recent state versions. Because COLE's Merkle file constructs a complete Merkle Hash Tree (MHT) over all historical state versions for the corresponding value file, pruning historical states would disrupt its ability to correctly construct the Merkle file and compute the root hash for the new compacted run during subsequent LSM-tree merge operations.} This makes it impossible for full archive nodes and pruned nodes to agree on a consistent tree index and root hash for blockchain states. Consequently, all blockchain full nodes in COLE must maintain all historical states, even those that are rarely queried.
        
		
		\item \textbf{Uniform Indexing of State Versions}: 
        \ce{COLE treats all state versions equally, neglecting to account for their disparate access frequencies. Since blockchain queries predominantly access the latest state values, this uniform approach introduces inefficiencies by unnecessarily expanding the search space from the current state to all historical versions during data queries.}
	\end{itemize}
	
	\rev{To address the aforementioned limitations of COLE while retaining the benefits of its column-oriented design and learned index, this paper proposes COLE$^+$, a novel system that supports chain reorganization and efficient state pruning through novel storage layouts and innovative Merkle index designs.}

    \rev{Specifically, to enable efficient chain reorganization of recent blocks, we propose a novel in-memory \emph{rewind-supported tree} ($\memindex$). It integrates the content-defined chunking (CDC) concept~\cite{xia2016fastcdc, muthitacharoen2001low} to ensure a deterministic structure, yielding a consistent root hash for state rewinds and appends that is independent of the update order. To prevent flushing all states to disk and the possible rebuilding of on-disk runs after an LSM-tree flush, we maintain two groups of in-memory $\memindex$ instances along with temporary hash lists acting as checkpoints for each LSM-tree run before flushing. These hash lists allow us to retrieve the original hashes for pre-existing on-disk runs (those present before the flush) and enable computing the new root hash as if no reorganization had occurred. For rare, deeper reorganizations extending to arbitrary on-disk levels, our approach anchors the rewind point at the latest checkpoint before the common ancestor block being shared with the canonical chain. Only LSM-tree runs altered by this rewind are rebuilt, and subsequent canonical blocks are appended normally.}

    \rev{To facilitate efficient state pruning, we adopt a two-level MHT structure that separates the latest and historical versions of each state. Specifically, the lower level uses a specially designed prunable \emph{version tree} to index the historical versions of each state. The upper level consists of a complete MHT that incorporates the latest value of each state along with the corresponding version tree from the lower level. This two-level design also streamlines the retrieval of the latest states by narrowing the search space. The version tree facilitates pruning through a purposefully designed CDC algorithm for index construction, which ensures a deterministic index structure even when many states have been pruned. During disk-level merge sort operations, our custom CDC enables the construction of a new merged version tree using only the left-most and right-most boundary paths from the two merging trees, thereby safely pruning all intermediate nodes between the two boundary paths.}
    
	
	
	We provide both theoretical and empirical analyses to validate the effectiveness of the proposed techniques. 
	\rev{Experimental results show that, with pruning enabled, COLE$^+$  achieves a storage size reduction of up to $16.7\times$ and $98.1\times$ compared to COLE and MPT, respectively. Furthermore, COLE$^+$ improves throughput by up to $1.3\times$ and $3.7\times$ over COLE and MPT, respectively.} These results demonstrate the potential of COLE$^+$ to significantly reduce storage requirements and enhance throughput performance while also supporting chain reorganization in practical blockchain deployments.
	
	The remainder of the paper is organized as follows. Section~\ref{sec:overview} provides an overview of the COLE$^+$ designs. Section~\ref{sec:rewind-tree} presents the RS-tree and details its mechanism for handling chain reorganization. Section~\ref{sec:prune-version-file} introduces the proposed version tree, followed by a description of COLE$^+$'s write and read operations in Section~\ref{sec:write-read}. Section~\ref{sec:experiments} reports experimental results. Finally, we discuss related work in Section~\ref{sec:related_work} and conclude the paper in Section~\ref{sec:conclusion}.
	
	\section{COLE$^+$ Overview}\label{sec:overview}

    This section presents an overview of COLE$^+$ with its novel features: chain reorganization, state pruning, and an improved storage layout. We start by giving a brief introduction to COLE's structure, followed by the key designs in COLE$^+$.

	\subsection{Preliminary: COLE}\label{sec:cole-limitations}
    \rev{COLE employs a log-structured merge-tree (LSM-tree) to efficiently handle frequent blockchain state updates, taking advantage of its write-optimized design. The structure consists of an in-memory level and multiple on-disk levels with progressively larger capacities. When a state is updated in a new block, COLE inserts the updated state along with its version (i.e., block height) into the in-memory Merkle B-tree (MB-tree)~\cite{li2006dynamic} (i.e., $L_0$ as shown in \cref{fig:cole-overview}) using a \emph{compound key} $\mathcal{K}=\langle addr, blk\rangle$, where $addr$ is the state address and $blk$ is the block height.} For example, in \cref{fig:cole-overview}, when block $i+1$ updates the state at address $k_3$, the compound pair $\langle k_3, i+1\rangle$ together with the new value $v_3'$ is inserted into $L_0$. \rev{When $L_0$ reaches a predefined maximum capacity, it is flushed to disk as a sorted run in the next level $L_1$. Merge operations may then occur recursively across subsequent levels as needed, until no level exceeds its capacity threshold. Upon block finalization, COLE computes a state digest to attest to blockchain state integrity. This digest is derived from a \textsf{root\_hash\_list}, which aggregates the root hashes of the in-memory MB-tree and the Merkle Hash Trees (MHTs)~\cite{merkle1989certified} associated with each on-disk run. The \textsf{root\_hash\_list} is updated once the flush and merge operations are committed.}

    Each on-disk level consists of multiple sorted runs, each comprising three files:
    \begin{itemize}
		\item \textbf{Value file} contains the sorted blockchain states in the form of compound key-value pairs. For instance, the first run $R_1$ in $L_1$ includes entries such as $\langle \langle k_1, i-1\rangle, v_1 \rangle, \langle \langle k_1, i\rangle, v_1' \rangle$, and $\langle \langle k_2, i\rangle, v_2 \rangle$ shown in \cref{fig:cole-overview}.
		\item \textbf{Index file} holds a series of $\epsilon$-bounded piecewise linear models, which help efficiently locate blockchain states within the value file. Given a model, a compound key's position in the value file is predicted as $p_{pred}$ that satisfies $|p_{pred}-p_{real}|\leq \epsilon$, where $p_{real}$ is the key's real location. By setting $\epsilon$ to half the page capacity, at most two file pages will be accessed per model during data retrieval (i.e., the page of $p_{pred}$ and either its preceding or succeeding page), thereby enhancing I/O efficiency.
		\item \textbf{Merkle file} stores a complete MHT constructed from the states in the value file for the purpose of data authentication. With the complete MHT, it is able to generate a proof to verify a given piece of data based on its position in the value file, thus facilitating efficient data provenance.
	\end{itemize}

    For data retrieval in COLE, the process follows a level-wise search approach. To locate the state value corresponding to address $addr_q$ with version $blk_q$, a search key $\mathcal{K}_q \gets \langle addr_q, blk_q\rangle$ is formed. The search traverses the in-memory MB-tree and the learned models in the disk levels, and stops upon finding a key $\mathcal{K}_r\gets \langle addr_r, blk_r\rangle$ such that $addr_r = addr_q$ and $blk_r\leq blk_q$, at which point the corresponding value is returned. Retrieving the latest state value is similar but utilizing a special search key $\langle addr_q, max\_int\rangle$, where $max\_int$ is the maximum block height.

	\subsection{Key Designs in COLE$^+$}
	
	\begin{figure}[t]
		\centering
		\includegraphics[width=.95\linewidth]{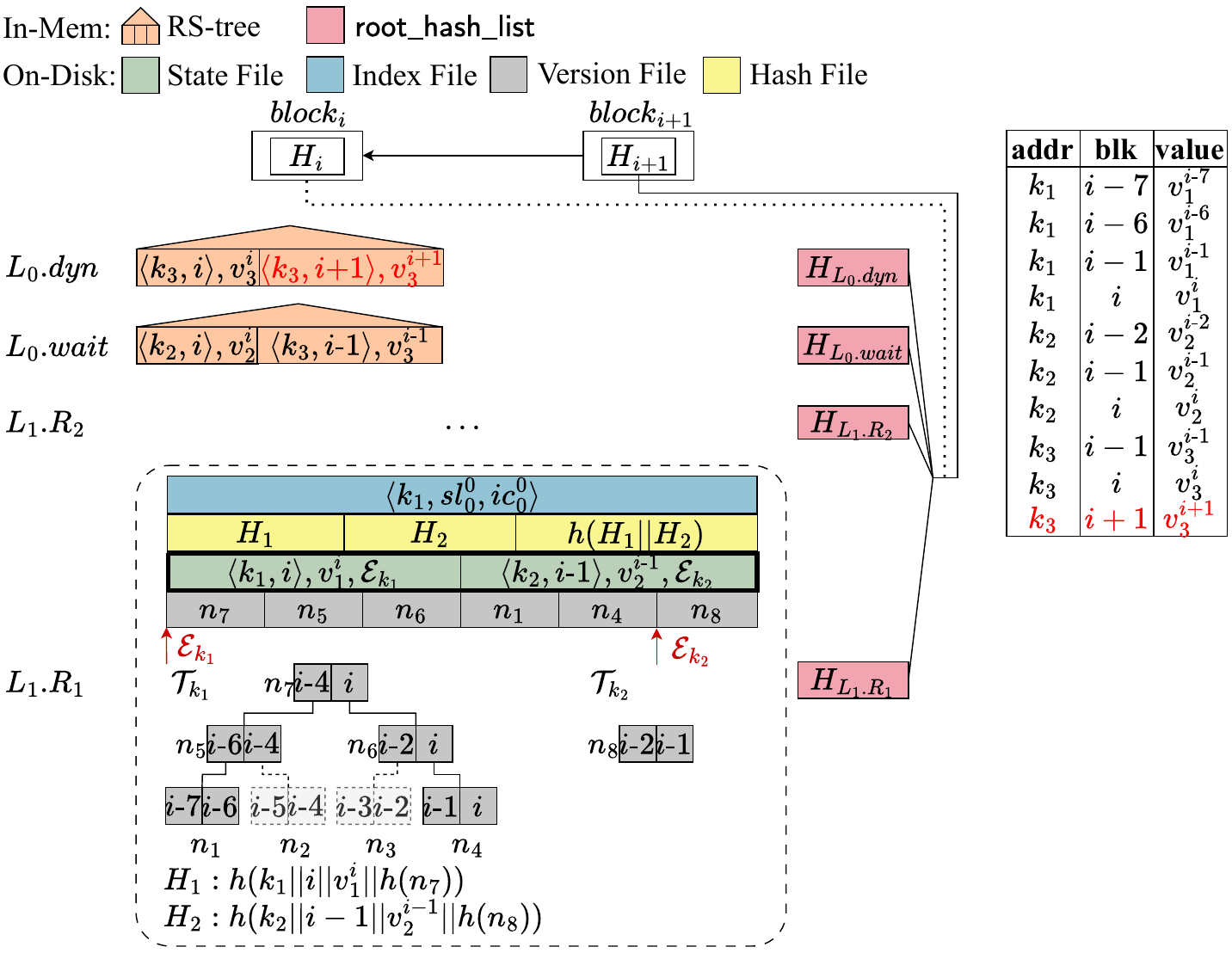}
        \vspace{-1em}
		\caption{Structure of COLE$^+$}\label{fig:cole-plus-overview}
	\end{figure}

    \rev{COLE$^+$ advances beyond COLE's limitations with several novel designs: rewind-supported trees for in-memory blockchain states, prunable version files for on-disk historical states, and an enhanced storage layer. The key designs of COLE$^+$ are detailed below.}
    
	
	\textbf{Rewind-Supported Tree}: 
    To support efficient chain reorganization occurring commonly for recent blocks as mentioned in \cref{sec:intro}, COLE$^+$ introduces a novel in-memory index structure, the rewind-supported tree ($\memindex$).
    Inspired by content-defined chunking (CDC)~\cite{xia2016fastcdc, muthitacharoen2001low}, the $\memindex$ determines node splitting points based solely on local data patterns, independent of the update sequence. This design ensures a deterministic tree structure, regardless of the removal of stale states and/or the appending of new states during chain reorganization. 
	The in-memory layer of COLE$^+$ consists of two groups of trees: the \emph{dynamic} group for incoming data and the \emph{waiting} group for data ready for the flush operation. This prevents the complete flushing of in-memory data, ensuring that enough states remain available for rewinding. During index rewind, if all rewind states belong to the dynamic group, they can be directly removed from the dynamic $\memindex$. If the rewind states span both groups, the current dynamic group is simply discarded, and the current waiting group is reverted back to the old dynamic group, with corresponding data being removed from its $\memindex$.
    \rev{The old waiting group is then restored from the flushed disk run, followed by the reverting of any possible merge operations on disk.} To avoid the costly disk operations in the second case, COLE$^+$ creates a temporary checkpoint by taking a snapshot of the \textsf{root\_hash\_list} before each flush operation.
    Although some states are flushed to disk, computing the index's root digest in the block only requires the updated in-memory $\memindex$ root hashes and the old snapshot hashes in the previous \textsf{root\_hash\_list} for on-disk runs. 
	
	\textbf{Prunable Version File}: The prunable version file is proposed for storing historical states on disk to support state pruning. It stores the historical values of each state along with their authenticated information, forming a \emph{version tree} for each state. The version tree is a specialized MHT that uses a purposely designed CDC algorithm for node construction. Because of the locality property of CDC, node splitting points are determined by the node's content rather than the update sequence, unlike a standard complete MHT. Therefore, LSM‐tree merge operations affect only the splitting points of a limited number of contiguous tree nodes bounded by the CDC algorithm. Consequently, all tree nodes outside the left-most and right-most boundary paths can be safely pruned. Importantly, regardless of pruning, the structure and corresponding root hash of the new version tree remain consistent after LSM-tree merge operations.
	For example, \cref{fig:cole-plus-overview} shows a version tree $\mathcal{T}_{k_1}$ for address $k_1$, where only the states' version numbers are displayed and the values are omitted. After state pruning, nodes $n_2$ and $n_3$ can be safely removed, while the nodes along the boundary paths (i.e., $n_7, n_5, n_6, n_1, n_4$) are retained for subsequent LSM-tree merge operations.


	
	\textbf{Improved Storage Layout}: COLE$^+$ further optimizes the storage layers for the on-disk level. Like COLE, it leverages the LSM-tree maintenance strategy for better write efficiency. However, for each disk run, COLE$^+$ utilizes four improved files: a \emph{state file}, an \emph{index file}, a \emph{version file}, and a \emph{hash file}. The state file records only the \emph{latest version} of each state in the run, along with pointers to its historical versions stored in the version file. 
    \rev{This separation of latest and historical versions improves the efficiency of data retrieval by reducing the search space.}
    As shown in \cref{fig:cole-plus-overview}, the state file of run $L_1.R_1$ contains the latest version of $k_1$, denoted as $v_1^i$, and the latest version of $k_2$, denoted as $v_2^{i-1}$, along with their respective pointers, $\mathcal{E}_{k_1}, \mathcal{E}_{k_2}$. The index file contains disk-optimized learned models that serve as an index for searching values in the state file, similar to the approach used in COLE\@. However, as we observe that the index must read entire pages whenever accessing data from the disk, COLE$^+$ chooses to reduce the precision of model predictions and the training input data size, 
    \ce{thereby improving the training efficiency.}
    \rev{The version file stores the historical versions of each state using a novel design. The hash file maintains a complete MHT over the latest states from the state file, and is associated with the root hash of the corresponding version tree.}
    To compute the index digest in the block header, the root hashes of the two $\memindexs$ in memory and the hash file on disk are stored in the \textsf{root\_hash\_list}. In \cref{fig:cole-plus-overview}, the hash file of run $L_1.R_1$ contains a complete MHT with two leaf hashes, $H_1$ and $H_2$, which authenticate the latest values and the version trees of addresses $k_1$ and $k_2$, respectively. The root hash is computed by hashing the concatenation of $H_1$ and $H_2$.
	
	\section{Rewind-Supported Tree}\label{sec:rewind-tree}
	This section introduces the in-memory rewind-supported tree ($\memindex$), which utilizes a CDC-based approach for node splitting. We start with a brief overview of the content‐defined chunking (CDC) algorithm, then show the construction of an $\memindex$, and finally explore the process of chain reorganization during a blockchain fork.
    
    \vspace{-0.3em}
	\subsection{CDC Algorithm}\label{sec:cdc}
    \vspace{-0.3em}
	
	A straightforward approach to node construction during tree building is splitting based on fanout. For example, in a B+-tree or a complete MHT, a node splits upon reaching its maximum fanout. However, this method suffers from a key drawback: node splitting points are sensitive to data updates. Inserting a new data entry at the beginning, for instance, shifts the entire existing data sequence, potentially invalidating almost all nodes due to altered splitting points. This non-deterministic index structure, dependent on both the update sequence and the data content, violates the requirement for a consistent root hash, especially during chain reorganization.
	
	\rev{To make the index structure independent of the data update sequence, we adopt the content-defined chunking (CDC) method~\cite{xia2016fastcdc, muthitacharoen2001low} to determine the node splitting points based solely on local data patterns. Owing to this locality property, the index structure is determined entirely by the indexed data itself. The CDC method consists of two stages: (i) computing a \emph{fingerprint} using a sliding window over the data content, and (ii) comparing the fingerprint against a \emph{mask} derived from the expected chunk size to decide whether to create a cut point (i.e., finding a CDC pattern). Several rolling-hash algorithms can be used to implement the fingerprint, such as Gear Hash~\cite{xia2014ddelta} and Rabin~\cite{broder1993some, rabin1981fingerprinting}.}
	
	\rev{To adapt the CDC method for COLE$^+$, we make the following essential modifications: (i) introducing a maximum chunk size $f_{max}$, which effectively limits the maximum fanout of each tree node, (ii) aligning cut points with the data entry size (256 bits for both state values in leaf nodes and hash values in internal nodes), and (iii) independently finding a tree node's cut point by resetting the fingerprint before examining the next one. The first modification prevents excessively large tree nodes. The second and third modifications are designed for the blockchain context. Unlike the original CDC method, which was designed for byte-stream deduplication, our proposed CDC algorithm ensures that tree node cut points are aligned with the 256-bit entry size. Moreover, cut points are determined solely by the data within the current node, rather than by data across multiple nodes, which is vital for state pruning.}

	\begin{algorithm}[t]
		\caption{CDC Algorithm in Tree Nodes}\label{alg:cdc-hash}
		\SetKwFunction{FInit}{InitParams}
		\SetKwBlock{slide}{slide each $|w|$ bytes in $data$}{end}
		\Fn{\FInit{$f_{exp}, f_{max}$}}{
			\KwIn{Expected fanout $f_{exp}$, maximum fanout $f_{max}$}
			\KwOut{Parameter $param_{cdc}$}
			$param_{cdc}.mask\gets generate\_mask(f_{exp})$\;
			$param_{cdc}.cnt\gets 0$; $param_{cdc}.f_{max}\gets f_{max}$\;
			\Return{$param_{cdc}$}\;
		}
		\SetKwFunction{FMain}{CutPoint}
		\Fn{\FMain{$param_{cdc}, data$}}{
			\KwIn{Parameter $param_{cdc}$, input data chunked in 256 bits $data$}
			\KwOut{Pattern result}
			\rev{\If{$param_{cdc}.cnt > param_{cdc}.f_{max}$}{\label{alg:cdc-hash-1}
				$param_{cdc}.cnt \gets 0$\;
				\Return{$\text{CUT}$}\;\label{alg:cdc-hash-2}
			}
            $h_{cdc}\gets init\_cdc()$; $w\gets init\_window()$\;\label{alg:cdc-hash-3}
            \slide{\label{alg:cdc-hash-4}
                $slice\gets$ data slice in window $|w|$\;
                $fp\gets h_{cdc}.fingerprint(slice)$\;
                \If{$fp \ \&\  param_{cdc}.mask = 0$}{
                    $param_{cdc}.cnt \gets 0$\;
                    \Return{$\text{CUT}$}\;
                }\label{alg:cdc-hash-5}
            }
            $param_{cdc}.cnt \gets param_{cdc}.cnt + 1$\;
            \Return{$\text{NOCUT}$}\;
			}
			
		}
	\end{algorithm}
    \rev{\Cref{alg:cdc-hash} shows our proposed CDC algorithm for COLE$^+$. The function $\FuncSty{InitParams}(\cdot)$ initializes a parameter object that includes a CDC mask, a current node size counter $cnt$, and a maximum node size $f_{max}$. The CDC mask is determined by the expected node size, calculated from $f_{exp}$ multiplied by the data entry size (e.g., state key size + version size + value size).}
    In the function $\FuncSty{CutPoint}(\cdot)$, when the counter $cnt$ reaches $f_{max}$, a cut point is returned (\crefrange{alg:cdc-hash-1}{alg:cdc-hash-2}). Otherwise, the $\FuncSty{CutPoint}(\cdot)$ function generates the CDC fingerprint using a sliding window over the input data to check for the CDC pattern (\crefrange{alg:cdc-hash-4}{alg:cdc-hash-5}). If the fingerprint shares the same least-significant bits as the mask (i.e., $fp\ \&\ mask = 0$), a pattern (or cut point) is identified. Note that the CDC rolling hash will be reinitialized to clear any boundary effects before conducting the pattern check (\cref{alg:cdc-hash-3}).
    
	\subsection{Structure and Maintenance of $\memindex$}\label{sec:rewind-tree-maintain}
	The $\memindex$ resembles the structure of Merkle B-tree (MB-tree)~\cite{li2006dynamic}, with each node identified by its hash value. A leaf node contains key‐value pairs, $\{\langle \mathcal{K}_i, value_i\rangle\}_{i=1}^m$, and its hash is computed as $h(\mathcal{K}_1||value_1||\cdots ||\mathcal{K}_m||value_m)$, where $h(\cdot)$ is a cryptographic hash function (e.g., SHA-256) and `$||$' denotes concatenation. A non‐leaf node contains the search keys for locating child nodes and the hashes of those children, $\{\langle \mathcal{K}_{c_i}, h_{c_i}\rangle \}_{i=1}^{m}$. The hash of a non-leaf node is computed as $h(\mathcal{K}_{c_1} || h_{c_1}||\cdots|| \mathcal{K}_{c_m} || h_{c_m})$. When a child node is updated, its hash changes, subsequently altering the parent node’s hash, propagating up to the root. The root node’s hash can attest to all indexed data in the leaf nodes. However, unlike the MB‐tree, which splits nodes upon reaching the maximum fanout, the $\memindex$ splits nodes based on a CDC fingerprint matching a specific pattern or reaching the maximum fanout. 
	
\begin{figure}[t]
		\centering
		\includegraphics[width=0.7\linewidth]{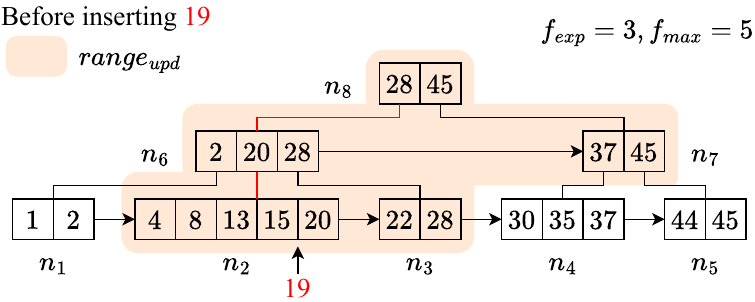}
        \vspace{-1em}
		\caption{$\memindex$ Before Inserting 19}\label{fig:rs-tree-before}
	\end{figure}

	\begin{example}
		\Cref{fig:rs-tree-before} shows an example of an $\memindex$. The maximum fanout is $f_{max}=5$. For simplicity, keys are used to represent node entries. Cut points are created for different reasons: nodes $n_1$, $n_3$, $n_4$, and $n_6$ are cut due to a matching CDC pattern on their last entries, while node $n_2$ reaches the maximum fanout. The remaining nodes are on the right boundary path. The hash of leaf node $n_2$ is $h(4||8||13||15||20)$. The hash of internal node $n_6$ is $h(2||h_{n_1}||20||h_{n_2}||28||h_{n_3})$.
	\end{example}

	\begin{algorithm}[t]
		\caption{$\memindex$ Maintenance (Insertion)}\label{alg:cdc-tree-insertion-sketch}
		\SetKwFunction{RSTreeInsert}{RSTreeInsert}
		\SetKwFunction{FCreate}{CDCCreateNodes}
		\SetKwFunction{FParent}{Par}
		\SetKwFunction{FSucc}{Succ}
		\Fn{\RSTreeInsert{$key, value$}}{
			\KwIn{Inserted key $key$, value $value$}
			$leaf\gets search\_key(key)$\;\label{alg:cdc-tree-insertion-sketch:1}
			$range_{upd}\gets[leaf, \FSucc(leaf)]$\;\label{alg:cdc-tree-insertion-sketch:2}
			\SetKwBlock{bTraverse}{Bottom-up traverse $\memindex$ along $key$'s path do}{end}
			\bTraverse{
				\lIf{at leaf level}{
					Update $\langle key, value\rangle$ in $range_{upd}$\label{alg:cdc-tree-insertion-sketch:3}
				}
				\Else{
					Replace obsolete entries in $range_{upd}$ with $e_{upd}$\;\label{alg:cdc-tree-insertion-sketch:4}
				}
				$nodes_{upd} \gets \FCreate(range_{upd}) $\;\label{alg:cdc-tree-insertion-sketch:5}
				Replace $range_{upd}$ in $\memindex$ with $nodes_{upd}$\;\label{alg:cdc-tree-insertion-sketch:6}
				$e_{upd}\gets \{ \langle \mathcal{K}_{c_m}^n, h(n)\rangle~|~\forall n\in nodes_{upd}\}$\;\label{alg:cdc-tree-insertion-sketch:7}
				$[head, tail] \gets range_{upd}$\;\label{alg:cdc-tree-insertion-sketch:8}
				$range_{upd}\gets [\FParent(head), \FSucc(\FParent(tail))]$\;\label{alg:cdc-tree-insertion-sketch:9}
			}
			If the root only has one entry, set its child as the new root\;\label{alg:cdc-tree-insertion-sketch:10}
		}
	\end{algorithm}
	
	The $\memindex$ insertion algorithm operates similarly to a traditional B+-tree, involving two traversals: a top-down traversal to locate the target leaf node and a bottom-up traversal to update nodes along the key’s path. However, unlike traditional B+-trees, which update at most one adjacent node per level, $\memindex$ may update multiple consecutive nodes at a given level. These additional updates stem from the CDC method, which can introduce multiple new cut points during updates. 
	To track these updates, we use a variable $range_{upd}$. When an entry is inserted into a node already at maximum fanout ($f_{max}$) or modifies a node’s last entry (its cut point), successive nodes are also affected and added to $range_{upd}$. At the leaf level, this process continues until a node with fewer than $f_{max}$ entries is encountered or the rightmost leaf is reached. Similarly, at non-leaf levels, $range_{upd}$ includes all updated child entries and extends to all necessary successive nodes.
	
	\Cref{alg:cdc-tree-insertion-sketch} outlines the $\memindex$ insertion procedure. First, it locates the target leaf node via a top-down traversal (\cref{alg:cdc-tree-insertion-sketch:1}). At the leaf level, the affected nodes $range_{upd}$ include the target leaf node and potentially its successive nodes (\cref{alg:cdc-tree-insertion-sketch:2}). Next, a bottom-up traversal is performed on the $\memindex$.
	At each level, data entries are first updated: the $\langle key, value \rangle$ pair is inserted into the corresponding leaf node (\cref{alg:cdc-tree-insertion-sketch:3}), while non-leaf nodes update their corresponding entries (\cref{alg:cdc-tree-insertion-sketch:4}). Then, all nodes in $range_{upd}$ are processed using the CDC method, generating a set of new nodes, $nodes_{upd}$, to replace the old ones (\crefrange{alg:cdc-tree-insertion-sketch:5}{alg:cdc-tree-insertion-sketch:6}).
	To maintain the search index structure, the corresponding search keys and hashes of these new nodes are collected into an entry list, $e_{upd}$, which will be used to update the corresponding parent nodes (\cref{alg:cdc-tree-insertion-sketch:7}). 
	The $range_{upd}$ for the parent level is then determined by the parent of the first node in $range_{upd}$ and the successive nodes of the parent of the last node in $range_{upd}$ (\crefrange{alg:cdc-tree-insertion-sketch:8}{alg:cdc-tree-insertion-sketch:9}). This bottom-up traversal continues until the root is reached. If the root contains only one entry (i.e., a single child), this child becomes the new root (\cref{alg:cdc-tree-insertion-sketch:10}).

	\begin{figure}[t]
		\centering
		\includegraphics[width=0.8\linewidth]{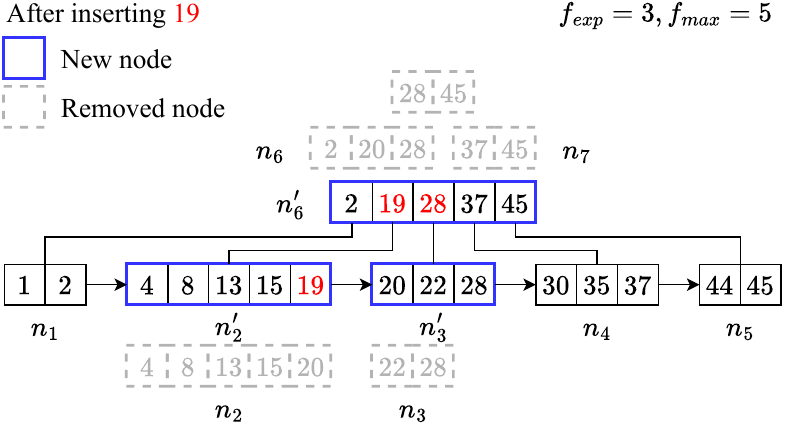}
        \vspace{-1em}
		\caption{$\memindex$ After Inserting 19}\label{fig:rs-tree-after}
	\end{figure}
	
	\begin{example}
		Figures \ref{fig:rs-tree-before} and \ref{fig:rs-tree-after} illustrate the insertion of a state with key 19, assuming a maximum fanout ($f_{max}$) of 5. First, the corresponding leaf node $n_2$ is located for the key $19$. The $range_{upd}$ at the leaf level includes both $n_2$ and its successive node $n_3$, which is included because $n_2$ has reached the maximum fanout. After inserting $19$ into $n_2$, the CDC algorithm creates new nodes $n_2'$ and $n_3'$, replacing their old counterparts $n_2$ and $n_3$. Next, their updated entries for the parent nodes, $\{\langle 19, h_{n_2'}\rangle, \langle 28, h_{n_3'} \rangle\}$, are added to the entry list, $e_{upd}$. The new $range_{upd}$ for the next level is computed as $[n_6, n_7]$ accordingly. We then process the next level. $\{\langle 20, h(n_2)\rangle, \langle 28, h(n_3)\rangle\}$ in node $n_6$ is replaced with $\{\langle 19, h_{n_2'}\rangle, \langle 28, h_{n_3'} \rangle\}$. After that, new cut points are generated by the CDC algorithm. Here, the original $n_6$ and $n_7$ are merged into a single node $n_6'$ due to pattern changes. The new entries $e_{upd}$ at this level are updated as $\{\langle 45, h(n_6')\rangle\}$, and the next $range_{upd}$ becomes $[n_8, n_8]$ as $n_8$ has no successive node. However, since $n_6'$ is the only node at the current level, it becomes the new root of the $\memindex$, and $n_8$ is discarded. In \cref{fig:rs-tree-after}, blue rectangles and dashed gray rectangles denote new nodes and removed nodes, respectively.

	\end{example} 
	
	The deletion operation is similar to the insertion, 
	with a key difference in how $range_{upd}$ is determined. If deleting an entry results in the removal of the last entry in a node, or if the node is at $f_{max}$  before deletion, successive nodes must be included in $range_{upd}$ to handle potential cut-point shifts.
    
	\subsection{Chain Reorganization in COLE$^+$}\label{sec:rewind}
    As mentioned in \cref{sec:intro}, chain reorganizations can occur either due to the eventual consistency of consensus protocols (e.g., Proof of Work in Bitcoin~\cite{nakamoto2008bitcoin} and Proof of Stake in Ethereum~\cite{wood2014ethereum}) or as a result of software upgrades (e.g., post-Ethereum's DAO attack~\cite{ethfork}). In consensus-related cases, temporary network partitions may lead to multiple concurrent chain branches before the network converges. Ultimately, only one branch becomes the canonical chain (e.g., Bitcoin's longest chain, which represents the majority of the network's computation power). During a chain reorganization, a  node on a non‐canonical chain first rewinds to the latest common ancestor block shared with the canonical chain. It then appends  and validates the new states from the canonical chain's remaining blocks by computing their canonical index digests. For less frequent software upgrades, the  process is similar but may require rewinding arbitrarily more blocks.
	
	
    We first focus on frequent consensus-related chain reorganizations, which involve only recent blocks.
    To support efficient state rewinding, we limit it to in-memory levels by maintaining two groups of $\memindexs$: a dynamic group and a waiting group. Both groups ensure the index’s root hash is determined solely by their content. New writes to COLE$^+$ are first inserted into the dynamic group. When the group reaches capacity, it is promoted to the waiting group, while the previous waiting group is flushed to the disk-level LSM-tree. Simultaneously, a new empty $\memindex$ initializes as the dynamic group. This design guarantees that state data is written to disk only after undergoing two flushes, which prevents premature flushing of all in-memory states and preserves sufficient in-memory states for future rewinds.
    
    If all non-canonical states requiring rewinding are contained within COLE$^+$'s dynamic group, they can be directly removed from the $\memindex$. Subsequently, new states from the canonical chain can be appended through normal write operations. Otherwise, if the rewind common ancestor block precedes the most recent flush operation, the current dynamic group is discarded and the current waiting group is reinstated as the dynamic group, with non-canonical states removed from its $\memindex$.
	Note that the last flush operation also modifies the disk levels. Although the on-disk runs could be rebuilt, doing so is computationally expensive and would undermine the requirement for rapid chain reorganizations in consensus protocols. \rev{Therefore, a new design is needed to enable appending states from the canonical chain without reverting on-disk changes previously caused by level merges.}
	COLE$^+$ creates a temporary checkpoint by snapshotting the \textsf{root\_hash\_list} before each flush operation. Although the current waiting group and all disk levels are out-of-sync with the nodes that do not undergo chain reorganization, computing the index digest (consequently validating the new block) only requires the current up-to-date dynamic group $\memindex$'s root hash and the previous hashes saved in the snapshotted \textsf{root\_hash\_list} for the out-of-sync waiting group and on-disk levels. Due to the inconsistency between the \textsf{root\_hash\_list} and the actual data on disk, index operations (e.g., read and write) are temporarily blocked. Once new states are appended to reach the point of the original flush operation, the waiting group and on-disk levels will catch up with the nodes without chain reorganization. After that, normal write operations are resumed.
	
\begin{figure}[t]
		\centering
		\includegraphics[width=\linewidth]{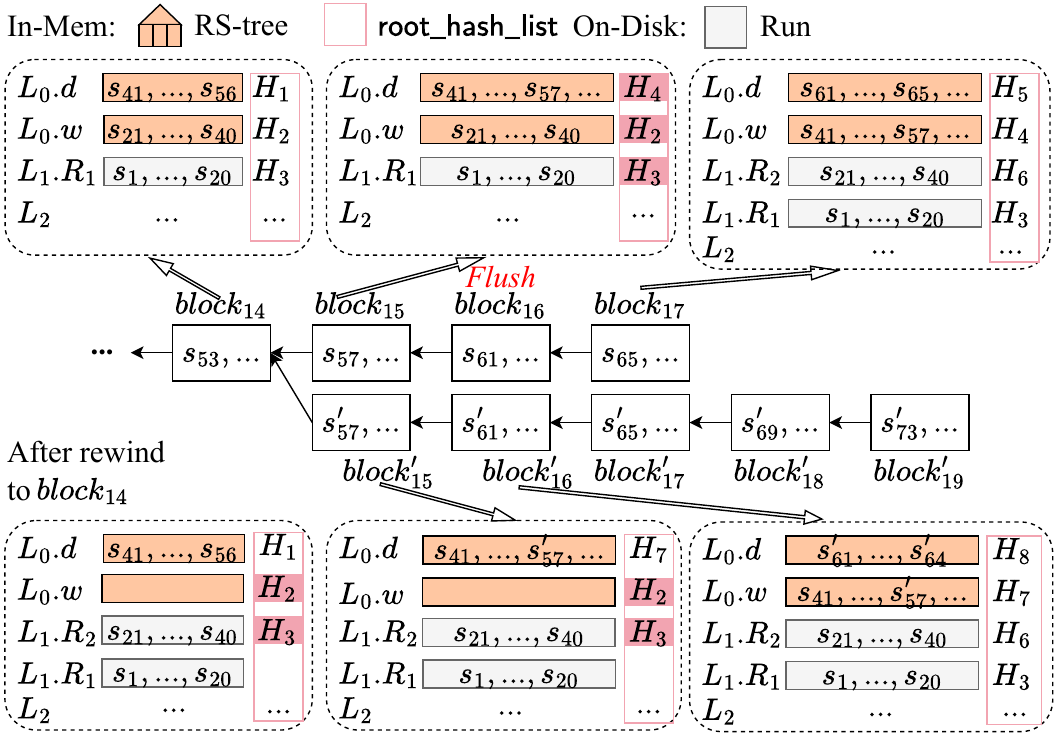}
        \vspace{-2em}
		\caption{Example of Chain Reorganization}\label{fig:fork}
	\end{figure}

	\begin{example}
        \rev{\Cref{fig:fork} shows how COLE$^+$ handles a chain reorganization. Assume each block updates four states, the in‐memory capacity is $40$, and $s_i$ denotes a state. At $block_{16}$, a flush occurs. The \textsf{root\_hash\_list} ($\{H_4, H_2, H_3, \dotsc\}$) is snapshotted. Then, the old dynamic group $L_0.d$ containing $s_{41}$ to $s_{60}$ becomes the new waiting group, while $s_{21}$ to $s_{40}$ in $L_0.w$ are flushed to $L_1.R_2$.
        Now consider a blockchain node at $block_{17}$ switching to $block_{19}'$. The node should first rewind states back to $block_{14}$ by removing states $s_{57}$ to $s_{68}$, then append the new states $s_{57}'$ to $s_{76}'$ from $block_{15}'$ to $block_{19}'$. Since the removed states span both groups, $s_{61}$ to $s_{68}$ in the dynamic group $L_0.d$ are discarded entirely. The old waiting group $L_0.w$ is reverted to the dynamic group, followed by removing $s_{57}$ to $s_{60}$ from its $\memindex$. At this point, the current dynamic group $L_0.d$ matches the original states of $block_{14}$ and yields the same hash value $H_1$. Although the waiting group and disk runs now differ from those in $block_{14}$, their old hash values are still available from the snapshotted \textsf{root\_hash\_list} (shaded in pink). Using these hash values $\{H_2, H_3, \dotsc\}$ and the up-to-date dynamic group $L_0.d$, the blockchain node computes the index digest for $block_{15}'$, even though the hashes of runs $L_1.R_2$ and $L_1.R_1$ temporarily differ from $H_2$ and $H_3$. Once both in-memory groups become full at $block_{16}'$, the entries in \textsf{root\_hash\_list} realign with both in-memory and on-disk levels. Since $L_1.R_2$ was flushed early, prior to $block_{15}'$, there is no need for an additional flush when both in-memory groups eventually fill up. The states $s_{61}'$ to $s_{76}'$ are then added through normal writes.}

	\end{example}
    
    \rev{Rare chain reorganizations, typically caused by software upgrades, necessitate extensive on-disk state rewinds. While recursive merges make rebuilding LSM-tree levels unavoidable, the checkpoint's \textsf{root\_hash\_list} minimizes the rebuilding of runs. The rewind point is defined as the most recent checkpoint immediately preceding the common ancestor block. Compared with its \textsf{root\_hash\_list}, matching runs are retained, missing runs are rebuilt from the current index states (their sorted nature facilitating consistency), and extra runs are discarded. Normal writes then resume to catch up with the canonical chain. The detailed algorithm is provided in
    \iftechreport%
    \cref{sec:reorg-disk}.
    \else
    our technical report \cite{techreport}.
    \fi
    }
    
    \rev{
    \textbf{Correctness Analysis}: The correctness of reorganization stems from the consistent root hashes maintained at each level. For frequent chain reorganizations, the in-memory RS-tree, leveraging CDC, ensures deterministic node cut points and index structure, thereby yielding a consistent root hash. For rare reorganizations, any changed on-disk runs are rebuilt. The sorted nature of these runs ensures deterministic index construction, thereby maintaining overall consistency. The detailed proof is provided in
    \iftechreport%
    \cref{sec:reorg-proof}.
    \else
    \cite{techreport}.
    \fi
    }
	\section{Prunable Version Tree}\label{sec:prune-version-file}
	This section describes the version tree, which is used to store the historical versions of a state within disk runs. We first detail the design of the version tree, including its state pruning capabilities.
	Then, we explain how the version trees of a given state, located in a level's multiple disk runs, are merged during an LSM-tree merge operation. 
	
	\subsection{Version Tree Structure}
	State pruning is a common technique used by blockchain full nodes to minimize storage overhead. Since historical versions of the state are rarely queried, blockchain full nodes can choose to retain only a few recent versions for each state. To support state pruning, in COLE$^+$, all historical versions of each state are stored as a version tree within each on-disk run. The structure of a version tree is similar to $\memindex$, where the CDC method is used to cut nodes at each level. Each leaf node contains version-value pairs $\{\langle blk_i, value_i \rangle\}_{i=1}^m$ with $h(blk_1||vaule_1||\dotsm||blk_m||value_m)$ as the node's hash. On the other hand, the non-leaf nodes contain the version number search key and corresponding hash for the child nodes, $\{\langle blk_{c_i}, h_{c_i} \rangle\}_{i=1}^m$ with the node's hash as $h({blk}_{c_1}|| h_{c_1}||\dotsm$$||{blk}_{c_m}$$||h_{c_m})$. Similarly, the hash of the root node is used to authenticate all version states stored in the leaf nodes. However, unlike $\memindex$ always being a full tree, the version tree can be either a full tree or a pruned tree.
	
    For pruned version trees, a key challenge arises: how to delete most tree nodes while still enabling the computation of the new merged version tree during subsequent LSM-tree merge operations. 
    Thanks to the locality property of the CDC method, the node cut points at each level are fully determined by the content of the tree nodes. This ensures that both full archive and pruned blockchain nodes can agree on the same version tree structure and, consequently, the same digest for the entire COLE$^+$ index. However, sufficient information needs to be retained such that pruned blockchain nodes can still compute updated nodes during the LSM-tree merge operation. We find that preserving the boundary paths (leftmost and rightmost) during state pruning could satisfy this requirement. Since the merged version tree follows a natural chronological order, the version spaces of the two trees are guaranteed not to overlap with each other. For example, within a level containing runs $R_i, R_{i-1}, \dotsc, R_{1}$ (ordered from newest to oldest), the versions in $R_i$ are guaranteed to be greater than those in $R_j$, where $i > j$. Due to this monotonic property, the merging process only requires the rightmost path of the left merged tree and the leftmost path of the right merged tree to determine node split points. All other nodes between these boundary paths can be safely discarded after pruning.
	
	Note that while only one node per level needs to be kept along the rightmost path of the left tree, retaining only the leftmost node at each level along the leftmost path of the right tree may be insufficient.
    \rev{If the leftmost node of the right tree has $f_{max}$ entries rather than following a CDC pattern, merging it with the rightmost node of the left tree might create a new node containing extra entries that do not satisfy a CDC pattern while having fewer entries than $f_{max}$. Therefore, at each level of the version tree, if the leftmost node has $f_{max}$ entries, all its successive nodes are retained until one with fewer than $f_{max}$ entries is encountered. All ancestor nodes of these boundary path nodes are likewise retained to maintain tree connectivity.}

	\begin{example}
		\Cref{fig:prune-file-example} shows two pruned version trees, $\mathcal{T}_l$ and $\mathcal{T}_r$, of one state, assuming the maximum fanout $f_{max}=3$.
        \ce{For brevity, only the version numbers of the states are shown.}
        For $\mathcal{T}_l$, the left-most and right-most nodes at each level are retained, while $\mathcal{T}_r$ retains all the nodes. Node $n_9$ cannot be pruned because its predecessor, $n_8$, is at $f_{max}$ capacity. There may be a situation, where after merging, versions $11$ and $12$ in $n_8$ cannot form a valid tree node on their own. Instead, a new node, $n_8'$, might need to be created using these extra entries along with version $13$ from $n_9$. As such, node $n_9$ must be retained.
		
	\end{example}
	
	\subsection{Merging Version Trees}\label{sec:merge-version-tree}
	
	\begin{figure}[t]
		\centering
		\includegraphics[width=0.8\linewidth]{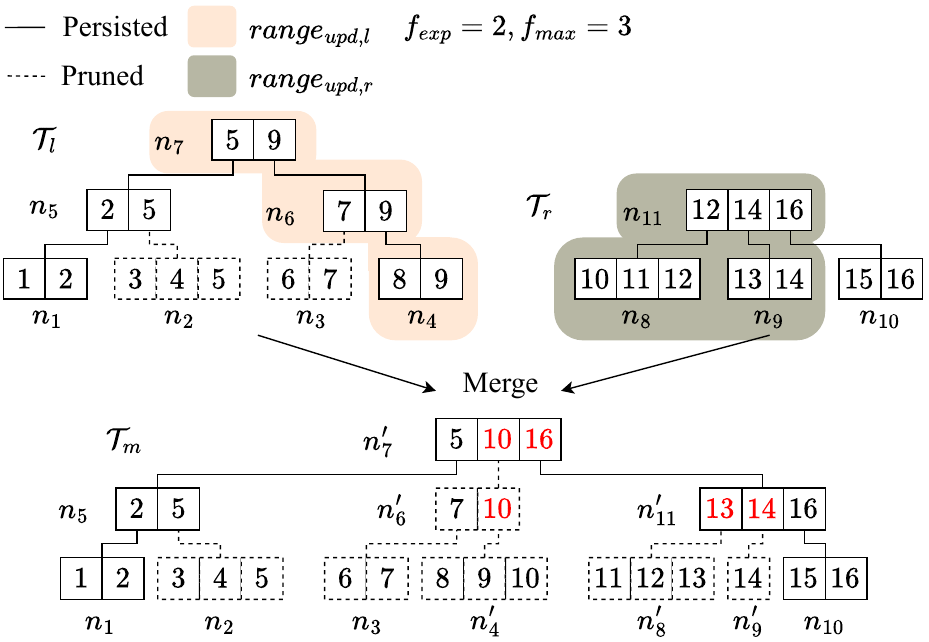}
        \vspace{-1em}
		\caption{Example of Version Trees}\label{fig:prune-file-example}
	\end{figure}
	
	\begin{algorithm}[t]
		\caption{Merge Version Trees}\label{alg:merge-version-tree}
		\SetKwFunction{FMergeTree}{MergeVersionTree}
		\SetKwFunction{FCreate}{CDCCreateNodes}
		\SetKwFunction{FParent}{Par}
		\SetKwFunction{FSucc}{Succ}
		\Fn{\FMergeTree{$\mathcal{T}_l, \mathcal{T}_r$}}{
			\KwIn{Left tree $\mathcal{T}_l$, right tree $\mathcal{T}_r$}
			\KwOut{Merged tree $\mathcal{T}_m$}
			$node_l \gets \mathcal{T}_l$'s rightmost leaf node\;\label{alg:merge-version-tree:1}
			$node_r \gets \mathcal{T}_r$'s leftmost leaf node\;\label{alg:merge-version-tree:2}
			$range_{upd,l}\gets[node_l, node_l]$\;\label{alg:merge-version-tree:3}
			$range_{upd,r}\gets[node_r, \FSucc(node_r)]$\;\label{alg:merge-version-tree:4}
			\While{$range_{upd,l} \neq \emptyset \lor range_{upd,r} \neq \emptyset$ }{
				Copy all nodes in the current level from $\mathcal{T}_l$, $\mathcal{T}_r$ to~$\mathcal{T}_m$\;\label{alg:merge-version-tree:5}
				$range_{upd,m} \gets range_{upd,l} \cup range_{upd,r}$\;\label{alg:merge-version-tree:6}
				\If{not at leaf level}{
					Replace obsolete entries in $range_{upd,m}$ with $e_{upd,m}$\;\label{alg:merge-version-tree:7}
				}
				$nodes_{upd} \gets \FCreate(range_{upd,m})$\;\label{alg:merge-version-tree:8}
				Replace $range_{upd,m}$ in $\mathcal{T}_m$ with $nodes_{upd}$\;\label{alg:merge-version-tree:9}
				$e_{upd}\gets \{ \langle {blk}_{c_m}^n, h(n)\rangle~|~\forall n\in nodes_{upd}\}$\;\label{alg:merge-version-tree:10}
				$[head_l, tail_l] \gets range_{upd,l}$\;\label{alg:merge-version-tree:11}
				$[head_r, tail_r] \gets range_{upd,r}$\;\label{alg:merge-version-tree:12}
				\lIf{$head_l$ is root node}{\label{alg:merge-version-tree:13}
					$range_{upd,l}\gets \emptyset$\label{alg:merge-version-tree:14}
				}\lElse{
					$range_{upd,l}\gets [\FParent(head_l), \FParent(tail_l)]$\label{alg:merge-version-tree:15}
				}
				\lIf{$head_r$ is root node}{\label{alg:merge-version-tree:16}
					$range_{upd,r}\gets \emptyset$\label{alg:merge-version-tree:17}
				}\Else{
					$range_{upd,r}\!\gets\![\FParent(head_r),\FSucc(\FParent(tail_r))]$\;\label{alg:merge-version-tree:18}
				}
			}
			If $\mathcal{T}_m$'s root only has one entry, set child as the new root\;\label{alg:merge-version-tree:19}
			\Return{$\mathcal{T}_m$}\;
		}
	\end{algorithm}

	\Cref{alg:merge-version-tree} details the procedure of merging two version trees, $\mathcal{T}_l$ and $\mathcal{T}_r$. It can be easily extended to support merging multiple trees by iteratively merging the resulting tree with the next one. Similar to \cref{alg:cdc-tree-insertion-sketch}, the procedure works by bottom-up traversing both $\mathcal{T}_l$ and $\mathcal{T}_r$ along the merging boundary paths. During the traversal, we maintain two ranges, $range_{upd,l}$ and ${range}_{upd,r}$, to track the new cut points from the CDC method for both left and right trees. Due to the aforementioned reason, at each tree level, $range_{upd,l}$ only contains the rightmost node of the left tree (\cref{alg:merge-version-tree:3,alg:merge-version-tree:15}), while ${range}_{upd,r}$ contains the leftmost node along with all necessary successive tree nodes for the right tree (\cref{alg:merge-version-tree:4,alg:merge-version-tree:18}). For each iteration of the traversal, we first copy all nodes at the current level from both trees to the merged tree (\cref{alg:merge-version-tree:5}). The update range of the merged tree $range_{upd,m}$ is computed by combining $range_{upd,l}$ and ${range}_{upd,r}$ (\cref{alg:merge-version-tree:6}). The updated entries $e_{upd}$ from the previous iteration are applied to this updated range, followed by creating new tree nodes using the CDC method (\crefrange{alg:merge-version-tree:7}{alg:merge-version-tree:8}). Next, the updated entries $e_{upd}$ are computed for the new tree nodes (\cref{alg:merge-version-tree:10}). Finally, we update ranges $range_{upd,l}$ and ${range}_{upd,r}$ for the next iteration (\crefrange{alg:merge-version-tree:13}{alg:merge-version-tree:18}). The tree traversal ends until reaching the tree roots for both $\mathcal{T}_l$ and $\mathcal{T}_r$. If the merged tree's root contains only one entry, this child becomes the new root (\cref{alg:merge-version-tree:19}).

	\begin{example}
		Following the example in \cref{fig:prune-file-example}, the merge operation is executed in a bottom-up fashion. Starting at the leaf level, $range_{upd,l}$ and $range_{upd,r}$ are computed as $[n_4, n_4]$ and $[n_8, n_9]$, respectively. Applying the CDC method, new nodes $n_4'$, $n_8'$, and $n_9'$ are created. These new nodes entail entry updates $e_{upd}$, consisting of $\{\langle 10, h(n_4')\rangle$, $\langle 13, h(n_8')\rangle$, $\langle 14, h(n_9')\rangle\}$. At the next level, $range_{upd,l}$ and $range_{upd,r}$ become $[n_6, n_6]$ and $[n_{11}, n_{11}]$, respectively. Consequently, new nodes $n_6'$ and $n_{11}'$ are generated with update entries $e_{upd} = \{\langle 10, h(n_6')\rangle$, $\langle 16, h(n_{11}')\rangle\}$. Finally, at the root level, $range_{upd,l}$ and $range_{upd,r}$ become $[n_7, n_7]$ and $\emptyset$, respectively. A root node $n_7'$ is created for the merged tree.
		The updated entries are highlighted in red. The merged tree can be further pruned, retaining only nodes, $n_7'$, $n_5$, $n_{11}'$, $n_1$, and~$n_{10}'$.
		
	\end{example}
	
    
    \rev{
    \textbf{Correctness Analysis}: The correctness of version trees stems from the consistent root hash achieved between full archive nodes and pruned nodes after merge operations.
    The version tree's structure is determined by the content-locality property of CDC. Full nodes trivially compute the correct root hash. For pruned nodes, any merge failure would imply the presence of essential updates outside the retained boundary paths, which directly contradicts CDC's locality property. Thus, retaining the  boundary paths ensures consistent root hashes across pruned nodes. The complete proof, as well as the analysis of the storage reduction, are given in
    \iftechreport%
    \cref{sec:analysis-merge-version-tree}
    and \cref{sec:analysis-storage-reduction}.
    \else
    \cite{techreport}.
    \fi
    }
    
	
	\section{Write and Read Operations in COLE$^+$}\label{sec:write-read}
    \rev{This section details the write and read operations in COLE$^+$.}
%
	\Cref{alg:cole-plus-write} outlines the write operation in COLE$^+$. The updated state's address $addr$ and the current block height $blk$ form the compound key $\mathcal{K}$. First, the compound key paired with the state value $value$ is inserted into the $\memindex$ of the dynamic group at level $L_0.dyn$ (\crefrange{alg:cole-plus-write-1}{alg:cole-plus-write-2}). When the dynamic group reaches half of the total in-memory capacity $\frac{B}{2}$, a flush operation begins. The current \textsf{root\_hash\_list} is stored as a snapshot to facilitate potential state rewinds (\cref{alg:cole-plus-write-5}). Then, the current waiting group $L_0.wait$ is flushed into an on-disk sorted run at $L_1$, creating four files: a state file $\mathcal{F}_S$, an index file $\mathcal{F}_I$, a version file $\mathcal{F}_V$, and a hash file $\mathcal{F}_H$ (\crefrange{alg:cole-plus-write-3}{alg:cole-plus-write-4}). Subsequently, the current dynamic group is promoted to the new waiting group, and a new empty waiting group is created for future write operations. Besides the in-memory flush operation, if any on-disk level $L_i$ reaches its capacity ($T$ runs), all runs in $L_i$ are merged into a new sorted run at level $L_{i+1}$ (\crefrange{alg:cole-plus-write-6}{alg:cole-plus-write-7}). During the process, multiple version trees for the same state address across different runs in $L_i$ are merged using the $\FuncSty{MergeVersionTree}(\cdot)$ function. Finally, the index digest $H_{index}$ for the new block is computed as $h(\textsf{root\_hash\_list})$, which captures the root hashes of $L_0$'s two $\memindexs$ and all on-disk runs (\cref{alg:cole-plus-write-8}).

    \begin{algorithm}[t]
		\caption{Write Algorithm}\label{alg:cole-plus-write}
		\SetKwFunction{FMain}{Put}
		\Fn{\FMain{$addr, value$}}{
			\KwIn{State address $addr$, value $value$}
			$blk\gets$ current block height; $\mathcal{K} \gets \langle addr, blk \rangle$\;\label{alg:cole-plus-write-1}
			Insert $\langle \mathcal{K}, value \rangle$ into the $\memindex$ in $L_0.dyn$\;\label{alg:cole-plus-write-2}
			\If{$L_0.dyn$ contains $\frac{B}{2}$ key-value pairs}{
				Store the temporary \textsf{root\_hash\_list}\;\label{alg:cole-plus-write-5}
				\If{$L_0.wait$ is not empty}{
					Flush the leaf nodes in $L_0.wait$ to $L_1$ as a sorted run\;\label{alg:cole-plus-write-3}
					Generate files $\mathcal{F}_S, \mathcal{F}_I, \mathcal{F}_V, \mathcal{F}_H$ for this run\;\label{alg:cole-plus-write-4}
					$L_0.wait.clear()$\;
				}
				Switch $L_0.dyn$ and $L_0.wait$\;
				$i\gets 1$\;
			}
			\While{$L_i$ contains $T$ runs}{\label{alg:cole-plus-write-6}
				Sort-merge all the runs in $L_i$ to $L_{i+1}$ as a new run\;
				Generate files $\mathcal{F}_S, \mathcal{F}_I, \mathcal{F}_V, \mathcal{F}_H$ for the new run\;
				Remove all the runs in $L_i$\;
				$i\gets i+1$\;\label{alg:cole-plus-write-7}
			}
			Update $H_{index}$ when finalizing the current block\;\label{alg:cole-plus-write-8}
		}
	\end{algorithm}
    
	Next, we detail the construction of the four files for each on-disk run. As mentioned in \cref{sec:overview}, the latest version of each state is stored in the state file, while historical versions are kept in the version file. This separation reduces the search space when querying the latest version, improving efficiency.
	The state file contains tuples of the form $\langle addr, blk, value, \mathcal{E}_{addr} \rangle$ for each state in the current run. The first three elements represent the compound key-value pair, while $\mathcal{E}_{addr}$ is a pointer to the corresponding version tree in the version file.
    \rev{Similar to COLE, a Bloom filter is built upon each $addr$ in a state file to expedite state searches.}
	The version file stores the version tree for each state by flattening its nodes in a breadth‐first search order.
	To authenticate both the state file and the version file, a hash file is created containing a complete Merkle Hash Tree (MHT). Each MHT leaf node is computed as $h(addr || blk || value || H_{\mathcal{T}_{addr}})$, where $value$ is the latest version in the state file and $H_{\mathcal{T}_{addr}}$ is the root hash of the corresponding version tree.
	The root hash of the hash file serves as the root hash of the current on-disk run and is stored in the \textsf{root\_hash\_list} for computing the final index digest. 
	
	Similar to COLE, the index file employs a hierarchy of piecewise linear models for efficiently indexing the state file. However, COLE$^+$ introduces several modifications:
	(i) the learned models use only the state address ($addr$) instead of the full compound key ($\mathcal{K}$) as input;
	(ii) model predictions return a page ID granularity rather than an exact offset in the state file;
	(iii) models are trained only on the first state in each page of the state file instead of all states; and
	(iv) the training error bound ($\epsilon$) for the piecewise linear models is set to 1.
	The first modification stems from COLE$^+$ relying on a separate version file for historical versions. The remaining changes are based on the observation that the index must read entire pages whenever accessing data from the disk, eliminating the need to predict high-precision offsets. These improvements decrease the size of training inputs and their precision requirements, leading to increased training efficiency.


	\begin{figure}[t]
		\centering
		\includegraphics[width=1\linewidth]{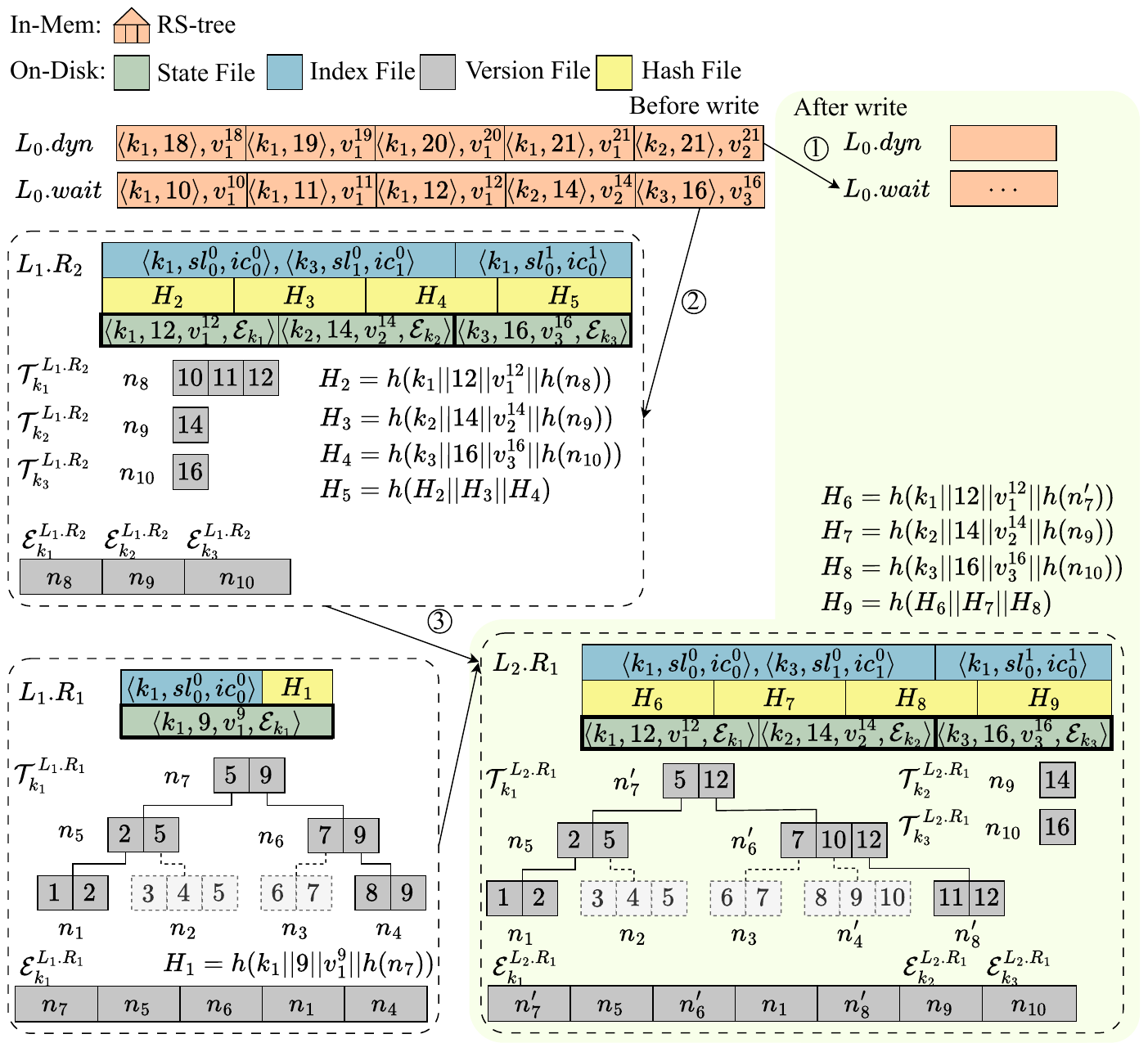}
        \vspace{-2em}
		\caption{An Example of Write Operation}\label{fig:write}
	\end{figure}
	
	\begin{example}
		\Cref{fig:write} illustrates the write operation in COLE$^+$. When $L_0.dyn$ reaches its capacity, \ding{172} it is promoted to $L_0.wait$. \ding{173} The previous $L_0.wait$ is then flushed to the disk as $L_1.R_2$. Note that the state file in $L_1.R_2$ only contains the latest entries for compound keys $\langle k_1, 12 \rangle$, $\langle k_2, 14 \rangle$, and $\langle k_3, 16 \rangle$. Since $L_1$ now contains two runs, \ding{174} they are merged into a new run at the next level $L_2.R_1$. As before, the state file at $L_2.R_1$ retains only the latest entries. The corresponding version file is constructed by merging $\mathcal{T}_{k_1}^{L_1.R_1}$ and $\mathcal{T}_{k_1}^{L_1.R_2}$ (via \cref{alg:merge-version-tree}) while copying $\mathcal{T}_{k_2}^{L_1.R_2}$ and $\mathcal{T}_{k_3}^{L_1.R_2}$. These version trees are stored on disk consecutively, with tree nodes arranged in breadth-first search order. The hash file's leaf nodes are computed as $H_6$, $H_7$, and $H_8$, with the root hash being computed as $H_9=h(H_6||H_7||H_8)$. This root hash serves as the digest for run $L_2.R_1$ by being added into \textsf{root\_hash\_list}. Finally, the index file is created by training piecewise learned models using inputs $\{\langle k_1, 0\rangle, \langle k_3, 1\rangle\}$, assuming a maximum of two entries per page in the state file.
	\end{example}
	
	There are two types of read operations in COLE$^+$, get queries and provenance queries. Get queries retrieve only the latest value for a given state $addr_q$.
	Like COLE, a get query in COLE$^+$ first searches the in-memory indexes, then on-disk runs in order of recency. However, the search keys differ between these levels. In the two $\memindexs$, a special key, $\mathcal{K}_q\gets\langle addr_q$, $max\_int\rangle$ is used to find the entry with the largest key $\mathcal{K}_r < \mathcal{K}_q$. If $\mathcal{K}_r.addr = addr_q$, its value is returned directly. Otherwise, the search proceeds to on-disk runs. 
    \rev{In contrast to COLE, the search key is simply $addr_q$, because the Bloom filters and the learned models are built directly on the state addresses. If the Bloom filter indicates the non-existence of $addr_q$, the state file is skipped. Otherwise, the learned model in the index file predicts the location within the state file using a page ID.} If the query entry is not in the corresponding page, either its preceding or succeeding page is retrieved and checked, limiting I/O to at most two pages per run. The search stops once the latest matching entry for $addr_q$ is found in the most recent run.
	
	\rev{The provenance query returns historical versions of the queried state $add_q$ within the block height range $[blk_l, blk_u]$, along with a Merkle proof rooted at the latest block to enable integrity verification.}
	First, the in-memory $\memindexs$ are searched using $[\langle addr_q, blk_l-1\rangle$, $\langle addr_q, blk_u+1\rangle]$, where the offsets ensure no versions are omitted from the results. During the search, the traversal path in the $\memindex$ is recorded as part of the Merkle proof. For on-disk runs, COLE$^+$ differs from COLE due to the separation of the latest and historical versions. First, the index file is queried with $addr$ to locate the corresponding version file offset $\mathcal{E}_{addr}$. If found, the version tree is searched using $[blk_l-1, blk_u+1]$, and the tree traversal path is added to the Merkle proof. Regardless of whether the query address appears in the current run, the corresponding Merkle path in the hash file is always included in the Merkle proof. The search stops once versions both earlier than $blk_l$ and later than $blk_u$ are encountered for $addr_q$. On the client side, verification involves recomputing the COLE$^+$ index root hash by reconstructing the Merkle tree using paths from the Merkle proof. The computed hash is then compared against the value stored in the block header.

	\begin{example}
		Following the example in \cref{fig:write}, where read operations occur after the write operation completes. Consider a get query of state $k_3$, $L_0$ is first searched using the compound key $\mathcal{K}_q\gets \langle k_3, max\_int \rangle$. $L_0.dyn$ returns nothing as it is empty. On the other hand, $L_0.wait$ returns entry for $\langle k_2, 21 \rangle$, which does not match $k_3$. Next, $L_2.R_1$ is searched as $L_1$ is also empty. Assuming the index file predicts that the address $k_3$ is located at the first page of the state file in $L_2.R_1$. In this case, the neighboring page (i.e., the second page) will be examined, yielding the final query result $v_3^{16}$.
		
		Next, consider a provenance query for state address $k_1$ within the version range $[10,18]$ using the non-pruned index. $[\langle k_1, 9\rangle, \langle k_1, 19\rangle]$ is used to search $L_0$, which returns empty and $\langle k_1, 18, v_1^{18}\rangle$ and $\langle k_1, 19, v_1^{19}\rangle$ from $L_0.dyn$ and $L_0.wait$, respectively. The corresponding Merkle path at the $\memindex$ is added to the Merkle proof. For on-disk runs, the index file allows us to locate $\mathcal{T}_{k_1}^{L_2.R_1}$ via pointer $\mathcal{E}_{k_1}$. At the same time, the Merkle path of the hash file, $\{H_7, H_8\}$, is added to the Merkle proof. To search the version tree, a version range $[9, 19]$ is used. The versions corresponding to block heights $9$, $10$, $11$, and $12$ are returned as part of the results, whereas $h(n_5)$, $h(n_3)$, $h(8||v_1^{8})$ are used for the Merkle proof. The client can reconstruct COLE$^+$'s index root hash to verify the soundness and completeness of the query results $\{v_1^{10}, v_1^{11}, v_1^{12}, v_1^{18}\}$.
	\end{example}

	\section{Experimental Evaluation}\label{sec:experiments}
	In this section, we compare COLE$^+$ with COLE \cite{zhang2024cole} and the Merkle Patricia Trie (MPT)~\cite{wood2014ethereum}, which serves as the index for the Ethereum blockchain. MPT is typically maintained by key-value databases such as RocksDB~\cite{rocksdb}. 
	\rev{Additionally, we evaluate the non-pruned COLE$^+$ (referred to as COLE$^+$-NP). LETUS~\cite{tian2024letus} is excluded from comparison because it is not open-source and demonstrates smaller performance gains over MPT than COLE~\cite{zhang2024cole}.} In the following subsections, we describe the system implementation and parameter settings, followed by the workloads and evaluation metrics. We then present the detailed experimental results.
    
	\vspace{-0.5em}
	\subsection{Implementation and Parameter Settings}
	COLE$^+$ is implemented in the Rust programming language with 16,000 lines of code 
    \cite{colepluscode}. 
    \rev{Both COLE$^+$ and COLE leverage asynchronous merge operations with multi-threading, introduced in \cite{zhang2024cole}, to mitigate tail latency during LSM-tree merges.} Blockchain transactions are executed using the Ethereum Virtual Machine (EVM). Each block processes 100 transactions, resulting in various read and write operations. Like COLE, COLE$^+$ uses simple file-based storage. The Gear Hash~\cite{xia2014ddelta} is employed to implement the CDC algorithm. 
    \rev{To maximize pruning efficiency, COLE$^+$ retains only the leftmost and rightmost boundary paths in each state's version tree, while pruning all intermediate nodes between these two boundary paths by default.}
    \Cref{tab:params} lists the parameters, with default values highlighted in bold. 
    \rev{The impact of parameters is evaluated in
    \iftechreport%
    \cref{sec:impact of parameters}.
    \else
    our technical report~\cite{techreport}.
    \fi
    The size ratio, indicating the maximum number of runs in a level, is set to 10 as it yields the highest throughput and comparably low latency. 
    The complete MHT's fanout is set to $4$ as it leads to the fastest provenance queries.
    }
	The experiments are conducted on a machine with an Intel i7‐10710U CPU 
	and a 1~TB SSD. 
	
	\begin{table}[t]
		\small
		\centering
        \caption{System Parameters}%
        \vspace{-0.5em}
		\begin{tabular}{ll}
			\toprule
			\textbf{Parameters}    & \textbf{Value}               \\ \midrule
			\# of generated blocks & $2\times 10^4, 6\times 10^4, 2\times 10^5, \mathbf{6\times 10^5}$ \\
			Size ratio $T$ & $2, 4, 6, 8, \mathbf{10}$ \\
			MHT fanout $f$     & $2, \mathbf{4}, 8, 16, 32$           \\
			\bottomrule
		\end{tabular}
		\label{tab:params}
	\end{table}
	\vspace{-0.5em}
    
	\subsection{Workloads and Evaluation Metrics}
	To simulate the blockchain workload, we use the KVStore workload from BlockBench~\cite{dinh2017blockbench} to generate blockchain transactions. Each transaction corresponds to a state read/update operation derived from the YCSB benchmark~\cite{cooper2010benchmarking}. Initially, $20,000$ transactions with new states are inserted into the storage as base data. Subsequently, various scenarios with different ratios of read/update operations are generated: (i) Write‐Only (entirely update operations); (ii) Read‐Write (half read and half update operations); and (iii) Read‐Only (entirely read operations). 
    \rev{Transactions are packed in blocks, which are committed serially using a single thread, as required by the blockchain's consensus protocol.}
    We measure storage size and system throughput to assess overall performance. For provenance queries, a set of state addresses is randomly selected from the base data, and different state version ranges extending from the latest block (e.g., $2, 4, \dotsc, 128$) are generated. We measure the total CPU time for both executing queries on the blockchain node and the query user’s verification time, as well as the proof size.
    \rev{We also evaluate block rewind latency for both frequent and rare cases as the number of rewound blocks increases. Additionally, we conduct an ablation study to assess the throughput impact of (i) the state separation layout (separating the latest states from historical ones), (ii) learned indexes for accelerating run searches, (iii) the improved CDC, and (iv) the proposed $\memindex$.
    }
    
	\subsection{Experimental Results}
	\subsubsection{Overall Performance}\label{sec:exp-overall-performance}
	\begin{figure}[t]
		\centering
		\includegraphics[width=.49\linewidth]{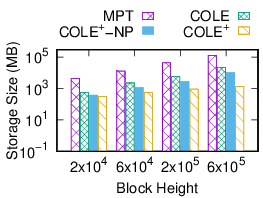}~%
		\includegraphics[width=.49\linewidth]{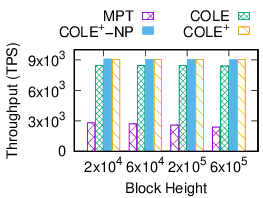}
        \vspace{-2em}
		\caption{Performance vs. Block Height}%
		\label{fig:write-only-uniform-perf}
	\end{figure}
    \Cref{fig:write-only-uniform-perf} compares the storage size and throughput of the evaluated indexes under the Write-Only workload.
    \rev{
    Compared to COLE, COLE$^+$-NP reduces the storage size by up to $2.2\times$ at a block height of $6\times 10^5$. This improvement stems from redesigning the state and version files. In COLE, each state address in a run is duplicated across historical versions, whereas COLE$^+$-NP stores each address only once in the run and relocates all historical values to the version file, while preserving the column-based design. State pruning in COLE$^+$ further reduces the storage size by $1.2\times$ to $7.7\times$ compared to COLE$^+$-NP, and by $16.7\times$ compared to COLE. This substantial gain arises from the innovative prunable version tree design, which retains only the boundary paths. Although COLE already achieves a $5.8\times$ storage reduction compared to MPT, COLE$^+$ achieves up to $98.1\times$ reduction over MPT.
    }
    
    \rev{
    Regarding system throughput, COLE$^+$-NP and COLE$^+$ achieve comparable throughput. While COLE already boosts throughput by up to $3.5\times$ over MPT thanks to its efficient column-based design, COLE$^+$ delivers an even higher improvement, up to $3.7\times$ over MPT, while also enabling state rewinds and pruning --- features that are critical for most practical blockchain applications.
    }

	\subsubsection{Impact of Workloads}
    \begin{figure}[t]
		\includegraphics[width=.49\linewidth]{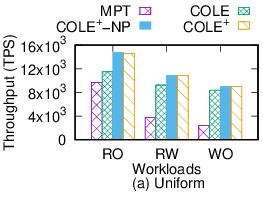}
		\includegraphics[width=.49\linewidth]{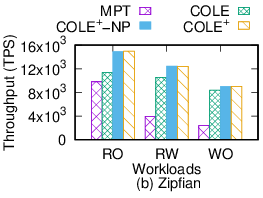}
        \vspace{-1em}
		\caption{Throughput vs. Workloads}\label{fig:workloads-uniform-zipfian-throughput}
	\end{figure}
	
	We use three workloads, Read-Only (RO), Read-Write (RW), and Write-Only (WO) across both Uniform and Zipfian distributions to assess their impact on system throughput. As shown in \cref{fig:workloads-uniform-zipfian-throughput}, all systems experience reduced throughput under increasing write operations. 
    \rev{Specifically, the throughput of MPT decreases by up to $75.5\%$. COLE's throughput decreases by up to $26.9\%$, while COLE$^+$-NP and COLE$^+$ experience comparable reductions, up to $39\%$ and $38.1\%$, respectively.} The LSM-tree-based maintenance approach used by COLE and COLE$^+$ generally enhances write performance. 
	
	Another interesting observation is that COLE$^+$ improves throughput over COLE under both Read-Only and Read-Write workloads. Under the Read-Only workload, COLE$^+$ achieves up to $28.7\%$ and $31.2\%$ higher throughput compared to COLE for the Uniform and Zipfian distributions, respectively. Under the Read-Write workload, COLE$^+$ also surpasses COLE, improving throughput by up to $17.7\%$. These performance gains are mainly attributed to COLE$^+$'s design that separates the latest and historical versions. Storing only the latest state in a state file reduces the search space for on-disk get queries, whereas COLE must search all historical versions.

	\subsubsection{Provenance Query Performance}
	\begin{figure}[t]
		\centering
		\includegraphics[width=.49\linewidth]{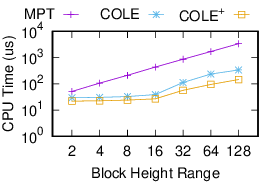}~%
		\includegraphics[width=.49\linewidth]{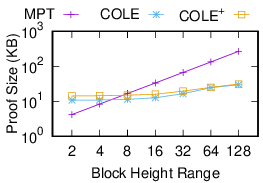}
        \vspace{-2em}
		\caption{Prov-Query Performance vs. Query Range}%
		\label{fig:prov}
	\end{figure}
	\rev{To evaluate provenance query performance, COLE$^+$ retains boundary paths that cover sufficient historical data (e.g., the most recent 128 blocks) instead of only the left-most and right-most boundary paths.} \Cref{fig:prov} compares the CPU time and proof size of MPT, COLE, and COLE$^+$ for provenance queries. For MPT, both metrics increase linearly with block height due to the requirement to query each block in the range. In contrast, COLE and COLE$^+$ exhibit sublinear growth. COLE$^+$ achieves superior CPU performance by separating latest state values from historical data, which enables its index models to filter non-queried states more effectively and thereby reduce the search space. The consecutive storage of version tree nodes further optimizes I/O costs. While COLE$^+$ has a slightly larger proof size than COLE, this increase results from the inclusion of version numbers as search keys in the version trees,  which are not required in COLE's complete MHT.
    \begin{figure}[t]
		\centering
		\includegraphics[width=.49\linewidth]{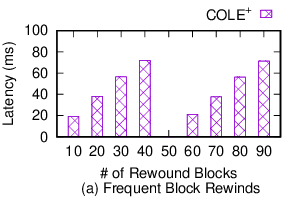}~%
		\includegraphics[width=.49\linewidth]{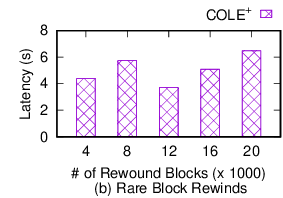}
        \vspace{-2em}
		\caption{Frequent and Rare Block Rewind Performance}%
		\label{fig:rewind-latency}
	\end{figure}
    
	\subsubsection{Block Rewind Performance}
	
	\rev{\Cref{fig:rewind-latency} shows the latency for frequent and rare block rewinds during chain reorganizations, across varying numbers of rewound blocks. Block appending results, similar to the Write-Only workload, are omitted. Frequent rewinds complete efficiently in under 100 ms. The drop at 50 blocks corresponds to the complete discarding of the dynamic group and only a few remaining removals from the waiting group. Rare rewinds are several hundred times slower, but still complete within 8 seconds for up to $2\times 10^4$ rewound blocks, which is acceptable in practice for infrequent events such as software upgrades. The drop at $12\times 10^3$ blocks stems from reusing unchanged on-disk runs, thus eliminating the costly run reconstruction process.}

    \begin{figure}[t]
		\centering
		\includegraphics[width=.6\linewidth]{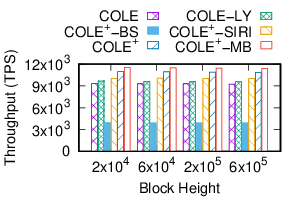}
        \vspace{-1em}
		\caption{Ablation Study}%
		\label{fig:ablation}
	\end{figure}
    
    \subsubsection{Ablation Study}
    \rev{We perform an ablation study with the following variants: 
    COLE-LY (adding latest/historical state separation to COLE),
    COLE$^+$-BS (replacing the learned model prediction with  binary search),
    COLE$^+$-SIRI (excluding the proposed CDC improvement),
    and COLE$^+$-MB (replacing COLE$^+$'s $\memindex$ with MB-tree~\cite{li2006dynamic}).
    As shown in \cref{fig:ablation}, COLE-LY increases throughput over COLE by 4\% under Read-Write workload. With more read operations, the new layout further boosts the gain to 14\%, as shown in 
    \iftechreport%
    \cref{sec:additional-ablation}.
    \else
    our techincal report \cite{techreport}, validating the effectiveness of the optimized storage layout.
    \fi
    COLE$^+$ achieves 2.8$\times$ throughput compared to COLE$^+$-BS, validating the effectiveness of the learned index. COLE$^+$'s improved CDC yields a 10\% increase in throughput compared to COLE$^+$-SIRI. Using $\memindex$ in COLE$^+$ incurs only a 5\% performance overhead compared to MB-tree, demonstrating an acceptable trade-off to enable efficient state rewind.
    }

    
	\section{Related Work} \label{sec:related_work}
	In this section, we provide a brief review of related studies on learned indexes and blockchain storage management.
	
	{\bf Learned Indexes.}
	Kraska \emph{et al.} introduced the concept of learned index structures by replacing search keys in an index node with a model, significantly reducing the search complexity and the size in terms of a node~\cite{kraska2018case}. Since then, numerous studies have been conducted to apply this approach across various scenarios. For example, \cite{ding2020alex, galakatos2019fiting, wu2021updatable, yu2022lifoss} focus on making learned indexes updatable. \cite{ferragina2020pgm} employs optimal piecewise linear models to construct indexes with theoretical worst-case bounds. Studies such as \cite{li2020lisa, nathan2020learning, tsunami20} address multidimensional data. \cite{zhang2024hyper} explores a hybrid construction to balance performance and memory consumption, while \cite{liang2024swix} targets memory-efficient sliding-window queries over data streams. 
    Beyond in-memory solutions, on-disk learned indexes, as discussed in \cite{zhang2024making, lan2024fully, lan2023updatable, dai2020wisckey}, account for the unique challenges of disk access and layout. 
    COLE~\cite{zhang2024cole} is the first learned storage for blockchain systems, supporting both data integrity and provenance queries. However, it fails to support chain reorganization and state pruning.
	
	{\bf Blockchain Storage Management.}
	Extensive studies have been proposed to optimize blockchain storage. Sharding technique is actively researched, where each node maintains only a partition of the blockchain, thereby reducing storage costs and enhancing parallelism~\cite{dang2019towards, el2019blockchaindb, zamani2018rapidchain, resilientdb20, jia2024estuary, xutwo}. Several studies have opted for off-chain storage solutions to alleviate on-chain costs~\cite{xu2021slimchain, honggridb23, teng2025timechain}. 
    ForkBase~\cite{wang2018forkbase} studied the problem of concurrent updates in a distributed network. It also employs the CDC method to efficiently identify and eliminate duplicate content across data objects in different branches to improve performance. 
    \rev{However, the CDC approach used by ForkBase is insufficient to directly support chain reorganization and state pruning.}
	Feng \emph{et al.} proposed SlimArchive, a storage optimization system for Ethereum full archive nodes. It reduces storage costs by flattening minimal blockchain state changes and removing Merkle-based structures~\cite{feng2024slimarchive}. However, this storage cost reduction comes at the cost of sacrificing data authentication and provenance, which are essential security features of blockchain systems.
	Tian \emph{et al.} proposed LETUS, a log-structured, trusted, universal blockchain storage system~\cite{tian2024letus}. LETUS employs a novel Merkle-based structure called DMM-Tree, which uses delta-encoding to reduce storage costs. However, similar to MPT, DMM-Tree retains obsolete nodes from historical blocks to support provenance queries.

	
	In addition to storage optimization, improving query efficiency in blockchain systems is also a promising research area. Some studies focused on verifiable queries over blockchain databases~\cite{wang2022vchain+, xu2019vchain, liu2023veffchain, wang2024v2fs}. Others explore query processing in the context of on-chain and off-chain hybrid storage~\cite{zhang2019gem, zhang2021authenticated, liu2024mpv, li2024authenticated, lin2024rollstore}. Recently, FlexIM~\cite{li2025flexim} has been proposed to manage and select verifiable indexes for dynamic queries in blockchain systems. Unlike these studies, COLE$^+$ and COLE focus on general-purpose blockchain storage.

	\section{Conclusion} \label{sec:conclusion}

	This paper has introduced COLE$^+$, an enhanced column‐based learned storage system for blockchains that addresses the limitations of existing state‐of‐the‐art solutions like COLE. By proposing a novel rewind‐supported in‐memory index structure based on CDC, COLE$^+$ enables efficient chain reorganization, a critical feature missing in COLE. Furthermore, COLE$^+$ introduces a new two-level MHT structure, incorporating a prunable version tree, that facilitates efficient state pruning and significantly reduces storage overhead. \rev{Empirical evaluations demonstrate that with pruning enabled, COLE$^+$ reduces storage size by up to $16.7\times$ and $98.1\times$ and increases throughput up to $1.3\times$ and $3.7\times$ compared to COLE and MPT, respectively.} These improvements, particularly the support for chain reorganization and state pruning, pave the way for wider practical adoption of COLE$^+$ by real-world blockchain systems. 
    \section{AI-Generated Content Acknowledgment}
    No AI tools were used to generate the text, figures, images, or pseudo-code in this paper.
	\bibliographystyle{IEEEtran}
	\bibliography{ref}
    \iftechreport\else\balance\fi
    \iftechreport
	\appendix
    \subsection{Glossary}
    \begin{table}[h]
    \centering
    \caption{Glossary}
    \small
    \begin{tabular}{lp{.65\linewidth}}
    \toprule
    \textbf{Term}                  & \textbf{Definition} \\ \midrule
    LSM-tree              &  Log-structure-merge tree          \\ 
    \textsf{root\_hash\_list} &  A list storing all the Merkle roots for both in-memory and on-disk levels       \\
    MB-tree               &  Merkle B-tree \\
    RS-tree               & Rewind-Supported tree           \\ 
    State File            &  Storing the latest versions of each state   \\
    Index File            &  Storing learned models for fast reads         \\
    Version File          &  Storing historical versions of each state    \\
    Hash File          &   Storing the complete Merkle Hash Tree built upon the version trees' root hash values         \\
    Checkpoint         &  A snapshot of the \textsf{root\_hash\_list} taken before the flush operation             \\
    CDC &  Content-defined chunking \\
    CDC Mask         &  A mask to control the average chunk size            \\
    \bottomrule
              
    \end{tabular}
    \end{table}

    \subsection{Discussion for Redesigned CDC}\label{sec:redesign-cdc}
    In \cref{sec:rewind-tree}, we have shown three essential modifications to the CDC design specifically for the blockchain context. The first modification involves establishing a maximum chunk size, $f_{max}$. This parameter limits the fanout of each tree node, which mitigates the risk of unbounded node growth. Such growth would inevitably compromise blockchain synchronization efficiency and impede the efficacy of state pruning. The second modification ensures that CDC cut points are consistently aligned with the blockchain's entry size. Blockchain data entries are characterized by their fixed size (256 bits for state and hash values) and atomic nature, meaning they cannot be subdivided. Arbitrary cut points, resulting from a lack of alignment, would violate this atomicity requirement and consequently degrade system performance. The third essential modification is to reset the CDC fingerprint at every cut point. This is paramount for COLE$^+$'s state pruning capabilities. Unlike traditional CDC algorithms that use information from preceding chunks to determine the next cut point, which is a valid approach when the entire data stream is always present, COLE$^+$ actively prunes historical states. Consequently, data prior to a cut point becomes inaccessible. To avoid dependencies on pruned data and ensure that COLE$^+$'s tree node boundaries are defined only by the contents of the current tree node, resetting the internal CDC fingerprint at each new node is a fundamental requirement.
    
	\subsection{Chain Reorganization with On-Disk Rewinds}\label{sec:reorg-disk}
	\begin{algorithm}[t]
		\caption{Chain Reorganization with On-Disk Rewinds}\label{alg:reorg}
		\SetKwFunction{FMain}{Chain-Reorg}
		\Fn{\FMain{$blk_{cur}, blk_{rew}, blk_{can}$}}{
			\KwIn{Current block $blk_{cur}$, rewound block $blk_{rew}$, latest canonical block $blk_{can}$}
			$\mathcal{L}_{cur}\gets$ Get \textsf{root\_hash\_list} for $blk_{cur}$\;\label{alg:reorg-1}
			$\mathcal{L}_{rew}\gets$ Get most recent \textsf{root\_hash\_list} happening before $blk_{rew}$\;\label{alg:reorg-2}
			$n_{cur}\gets |\mathcal{L}_{cur}|$; $n_{rew}\gets |\mathcal{L}_{rew}|$; $idx\gets 0$\;
			\tcc{Identify the unchanged runs via hash values}
			\ForEach{$\langle h_{cur}, h_{rew} \rangle $ \textbf{in} $\mathcal{L}_{cur}.rev(), \mathcal{L}_{rew}.rev()$}{\label{alg:reorg-3}
				\lIf{$h_{cur} \neq h_{rew}$}{
					\textbf{break}
				}
				$idx\gets idx + 1$\;
			}\label{alg:reorg-4}
			\tcc{Keep the unchanged disk runs}
			\For{$h_{cur}\in \mathcal{L}_{cur}[n_{cur}-idx, n_{cur}-1]$}{
				Retain the disk run w.r.t. $h_{cur}$\;
			}
			\tcc{Rebuild the inconsistent disk runs and $\memindex$}
			\For{$h_{rew}\in \mathcal{L}_{rew}[2, n_{rew}-idx-1].rev()$}{\label{alg:reorg-5}
				$\langle l_{rew}, r_{rew}\rangle\gets$ $h_{rew}.meta()$\;
				\eIf{$h_{rew}$ corresponds to a disk run}{
					Rebuild $h_{rew}$'s run using $\{s_{i}| s_{i}\in h_{cur}$'s run s.t.
					$s_{i}.blk\in \langle l_{rew}, r_{rew}\rangle, h_{cur}\in \mathcal{L}_{rew}\}$;
				}{
					Rebuild $h_{rew}$'s $\memindex$ using $\{s_{i}| s_{i}\in h_{cur}$'s run s.t.
					$s_{i}.blk\in \langle l_{rew}, r_{rew}\rangle, h_{cur}\in \mathcal{L}_{rew}\}$;
				}
			}\label{alg:reorg-6}
                Discard runs whose hash values are not in $\mathcal{L}_{rew}$\;
			\tcc{Append the remaining blocks of the canonical chain}
			\For{$blk \in$ sub-chain from $\mathcal{L}_{rew}$'s block to $blk_{can}$}{
				Execute TXs in $blk$\;
			}
		}
	\end{algorithm}
	In \cref{sec:rewind}, our primary focus is on chain reorganization at the in-memory level. Thanks to the careful design of temporary checkpoints, only the in-memory $\memindexs$ need to be rewound, while the on-disk runs remain unchanged, resulting in highly efficient chain reorganization. This type of reorganization occurs frequently due to the eventual consistency consensus protocols like Proof of Work in Bitcoin~\cite{nakamoto2008bitcoin} and Proof of Stake in Ethereum~\cite{wood2014ethereum}. However, there are cases where the state rewind extends beyond the in-memory level, such as during Ethereum's fork following the DAO attack~\cite{ethfork}, where lots of blocks are rewound. In such scenarios, simply rewinding the $\memindexs$ is insufficient because the on-disk runs have changed due to flush and recursive merge operations in the LSM-tree. To address this challenge, additional steps are necessary. The rewind point is tied to the most recent checkpoint prior to the common ancestor block shared with the canonical chain, as runs are only updated at these checkpoints. The changed on-disk runs are rebuilt using blockchain states stored in the current version index. Note that no transaction re-execution is required for rebuilding the on-disk runs, since the necessary data already exists, albeit distributed across different on‐disk runs. To further facilitate this rebuilding process, each checkpoint stores not only the root hash value of each run, but also metadata that includes the minimum and maximum block heights of the run’s updated states.
	
	\Cref{alg:reorg} shows the procedure of chain reorganization for on-disk levels. It takes three inputs: the current block $blk_{cur}$, the rewound block $blk_{rew}$ (the most recent common ancestor shared by the current and canonical chains), and the latest canonical block $blk_{can}$. First, both the snapshot corresponding to $blk_{cur}$ and the most recent snapshot that happens before $blk_{rew}$ are retrieved. (\crefrange{alg:reorg-1}{alg:reorg-2}). Next, to determine which runs should be retained, the common hash values between $\mathcal{L}_{cur}$ and $\mathcal{L}_{rew}$ are identified by iterating each list from the end (\crefrange{alg:reorg-3}{alg:reorg-4}). This reverse iteration is motivated by the LSM-tree merge sequence, where newer runs are more likely to have changed than older ones. The index, $idx$, records the number of these matching hashes. Any disk runs associated with these common hash values remain unchanged during the state rewind. For the remaining (changed) hash values in $\mathcal{L}_{rew}$, their corresponding disk runs or $\memindex$ are rebuilt (\crefrange{alg:reorg-5}{alg:reorg-6}).
	Specifically, the minimum and maximum block heights of the rebuilt run or $\memindex$ (i.e., $l_{rew}, r_{rew}$) are obtained from the metadata. The rebuild then uses the states in the current index whose update versions fall in $\langle l_{rew}, r_{rew}\rangle$. Runs whose hash values are absent from $\mathcal{L}_{rew}$ are then discarded. Finally, the remaining blocks of the canonical chain are appended, which is similar to the chain reorganization for the in-memory level.

    \subsection{Discussion for Adversarial Issues and ACID Properties}\label{sec:crash}
    COLE$^+$ focuses solely on the blockchain storage layer. Although blockchain systems as a whole are designed for adversarial environments, the adversarial issues are already addressed by the consensus protocol and cryptographic primitives. The storage layer of a blockchain system only needs to support basic data reads and writes while ensuring traditional ACID properties. 
    
    The atomicity is achieved by maintaining \textsf{root\_hash\_list} in an atomic manner. During the level merges, \textsf{root\_hash\_list} is updated atomically only after the completion of constructing all the files in the new level, followed by removing the old level files. This ensures data consistency because the old level files remain intact and are referenced by \textsf{root\_hash\_list} during a node crash. Concurrency control is not considered because the consensus protocol ensures the write serializability. The Merkle-based structures guarantee data integrity. For durability, COLE$^+$ utilizes checkpoints and blockchain transaction logs as Write-Ahead Logs to prevent data loss. 
    Checkpoints are created each time the in-memory index flushes to the disk. At this moment, only the waiting group is full, while the dynamic group is empty, and all other levels have already been persisted to disk. In the event of a crash at any point, the system can revert to the last checkpoint and safely discard all files associated with unfinished flush or merge operations. To reconstruct the in-memory waiting group, the system replays transactions occurring between the last two checkpoints. Similarly, to reconstruct the dynamic group created after the checkpoint, it replays additional transactions starting from the checkpoint up to the latest block.
    If a crash occurs during in-memory chain reorganization, we halt rewinding the in-memory RS-tree (since its state will be lost upon restart). Instead, the system treats this as a disk-level chain reorganization. In either case, a crash forces restoration to the most recent checkpoint preceding the rewind point, from which chain reorganization is restarted.
    
    \subsection{Proof of Correctness for Chain Reorganization}\label{sec:reorg-proof}
    \begin{theorem}[Chain Reorganization Correctness]\label{thm:rs-tree}
		COLE$^+$ supports chain reorganization with both in-memory rewind and on-disk rewind, and ensures consistent index digest between nodes with and without chain reorganization.
	\end{theorem}
	
	\begin{myproof}
		The index digest is computed from the hash values in the \textsf{root\_hash\_list}. To ensure its consistency, we analyze each entry in the list. For the dynamic group, it is guaranteed to have the same hash owing to the fact that $\memindex$'s hash depends completely on its stored data. For the waiting group and all on-disk runs, we prove by contradiction. Assume there is any inconsistency, it would imply there are some write operations in the waiting group or on-disk runs. This is impossible if the entire rewind happens at the dynamic group as all other LSM-tree runs remain unchanged during the time span. 
		If the rewind occurs in the waiting group, an observed inconsistency would imply that multiple flush operations occurred during the rewind, altering the snapshot hash values in the \textsf{root\_hash\_list}. This contradicts the assumption that the rewind happens solely in the waiting group. Alternatively, if the rewind extends to on-disk levels, the LSM-tree runs are rebuilt using the blockchain states in the existing index. An inconsistency would indicate additional merges occurred between snapshots, which is impossible since snapshots are taken before each flush operation.
		
	\end{myproof}

    \subsection{Proof of Correctness for Version Tree}\label{sec:analysis-merge-version-tree}

    \begin{theorem}[Version Tree Correctness]
		The proposed version tree ensures a consistent root hash between full archive nodes and pruned nodes after merge operations.
	\end{theorem}
	
	\begin{myproof}
		Since the structure of the version tree is fully determined by its content, different blockchain nodes should derive the same root hash value, provided they have sufficient information to construct the merged version tree. This is trivial for full archive nodes. For pruned nodes, we prove by contradiction that retaining tree nodes along the boundary paths is sufficient to construct the merged version tree. Assume, for contradiction, that pruned nodes cannot compute certain tree nodes during merging. This implies that updates occurred outside the boundary paths, beyond the retained nodes and the bounds established by the CDC pattern. Clearly, this is impossible as it violates CDC method's locality property.
	\end{myproof}
    
    \subsection{Complexity Analysis of Storage Reduction through State Pruning}\label{sec:analysis-storage-reduction}

    We finally analyze the storage reduction achieved by state pruning. Assume the number of versions for an address, denoted as $\mathcal{V}(addr)$, follows a Zipfian distribution, meaning a small number of addresses hold most historical versions. In COLE$^+$, state pruning operates on each address’s version tree. Therefore, we focus on the address with the most amount of historical versions. The size of a version tree is given by $O\left(f_{exp}\times (|k|+|v|) \times \frac{\mathcal{V}(addr)}{f_{exp} - 1} \right)$, where $f_{exp}$ is the expected fanout, $|k|$ is the size of the search key (i.e., version number), and $|v|$ is the size of the entry value (state value for leaf nodes or hash value for internal nodes). The height of the version tree is $\ceil{\log_{f_{exp}}\mathcal{V}(addr)}$. After state pruning, only the boundary paths are retained. The storage saving from pruning is: 
    \begin{equation*}
  \resizebox{0.98\linewidth}{!}{%
    $O\left(f_{exp}\times (|k|+|v|) \times \left(\frac{\mathcal{V}(addr)}{f_{exp} - 1}-2\times \ceil{\log_{f_{exp}} \mathcal{V}(addr)}+1\right) \right)$
  }
\end{equation*}
	For a large number of historical versions ($\mathcal{V}$ is large), the difference between $\frac{\mathcal{V}(addr)}{f_{exp} - 1}$ and $2\times \ceil{\log_{f_{exp}} \mathcal{V}(addr)}$ becomes asymptotically significant. This demonstrates the potential storage savings of state pruning while maintaining correctness.

    \subsection{Additional Experiment: Impact of Parameters}\label{sec:impact of parameters}

	\begin{figure}[h]
		\centering
		\includegraphics[width=.49\linewidth]{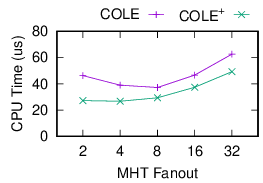}~%
		\includegraphics[width=.49\linewidth]{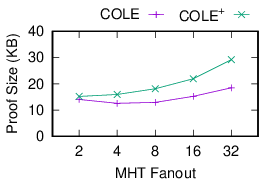}
        \vspace{-1em}
		\caption{Impact of MHT Fanout}%
		\label{fig:params-tuning-fanout}
	\end{figure}
	To evaluate the impact of fanout on CPU time and proof size for provenance queries, we vary the fanout exponentially from $2$ to $32$, while maintaining a fixed block height range of $16$. As shown in \cref{fig:params-tuning-fanout}, the CPU time exhibits a U-shaped trend, while the proof size of COLE$^+$ increases with larger fanouts. This occurs because larger fanouts reduce tree height, improving search efficiency, but also result in larger node sizes, increasing the proof size. The optimal fanout is $8$ for COLE and $4$ for COLE$^+$. We set the default fanout to $4$ in subsequent experiments.
	
	\begin{figure}[h]
		\centering
		\includegraphics[width=.49\linewidth]{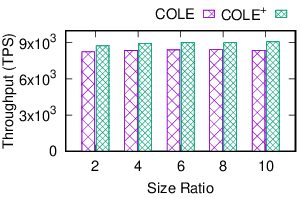}~%
		\includegraphics[width=.49\linewidth]{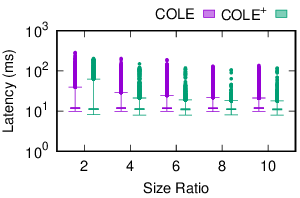}
        \vspace{-1em}
		\caption{Impact of Size Ratio}%
		\label{fig:params-tuning-size-ratio}
	\end{figure}
	\Cref{fig:params-tuning-size-ratio} compares the throughput and latency of COLE and COLE$^+$ for various size ratios $T$ under a block height of $6\times 10^5$. The throughput increases slightly with larger size ratios, while the latency decreases for both COLE and COLE$^+$. We set the default size ratio to $10$, as it yields the highest throughput and relatively low latency for COLE$^+$.

    \subsection{Additional Experiment: More on Ablation Study}\label{sec:additional-ablation}

    \begin{figure}[h]
		\centering
		\includegraphics[width=.49\linewidth]{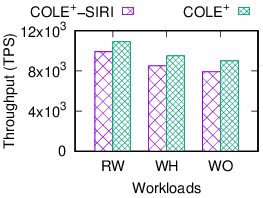}~%
        \vspace{-1em}
		\caption{COLE$^+$-SIRI Performance vs. Workloads}%
		\label{fig:siri-throughput}
	\end{figure}

    To evaluate the effectiveness of $\memindex$ versus SIRI~\cite{wang2018forkbase}, we test under varying workloads: (i) Read-Write (RW, 50\% update, 50\% read); (ii) Write-Heavy (WH, 75\% update, 25\% read); and (iii) Write-Only (WO, 100\% update). As shown in \cref{fig:siri-throughput}, COLE$^+$'s performance advantage increases with higher write intensity, rising from 10\% to 14\%, demonstrating the robustness of the RS-tree. Furthermore, it is worth noting that the CDC used in SIRI cannot directly support chain reorganization and state pruning, as discussed in \cref{sec:redesign-cdc}.

    \begin{figure}[h]
		\centering
		\includegraphics[width=.49\linewidth]{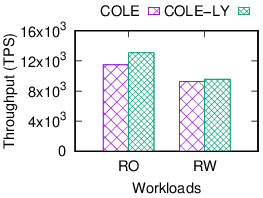}~%
        \vspace{-1em}
		\caption{COLE-LY Performance vs. Workloads}%
		\label{fig:layer-throughput}
	\end{figure}
    
    To assess the impact of the design of separating the latest and historical states, we compare baseline COLE with a variant of COLE featuring the same separated layout, denoted as COLE-LY.
    As shown in \cref{fig:layer-throughput}, COLE-LY improves throughput by up to 14\% over COLE under varying workloads. Thanks to the new layout, the search space of the latest states is reduced, thereby enhancing overall system throughput.



    \fi 

\end{document}
 